\documentclass[aps,prl,preprint,superscriptaddress]{revtex4-1}

\usepackage{graphicx}
\usepackage{epstopdf}
\usepackage{dcolumn}
\usepackage{bm}
\usepackage{color}
\usepackage{amsfonts}
\usepackage{amsmath}  
\usepackage{xcolor}
\usepackage[colorlinks=true,citecolor=blue]{hyperref}
\hypersetup{colorlinks=true,citecolor=blue,linkcolor=blue,urlcolor=blue}

\begin{document}


\title{In situ spacecraft observations of a structured electron diffusion region during magnetopause reconnection}




\author{Giulia Cozzani}
\email{giulia.cozzani@lpp.polytechnique.fr}
\affiliation{Laboratoire de Physique des Plasmas,CNRS/Ecole Polytechnique/Sorbonne Universit\'e, Universit\'e Paris Sud, Observatoire de Paris, Paris, France}
\affiliation{Dipartimento di Fisica ''E. Fermi'', Universit\`a Pisa, Pisa, Italy}
\author{A. Retin\`o}
\affiliation{Laboratoire de Physique des Plasmas,CNRS/Ecole Polytechnique/Sorbonne Universit\'e, Universit\'e Paris Sud, Observatoire de Paris, Paris, France}
\author{F. Califano}
\affiliation{Dipartimento di Fisica ''E. Fermi'', Universit\`a Pisa, Pisa, Italy}
\author{A. Alexandrova}
\affiliation{Laboratoire de Physique des Plasmas,CNRS/Ecole Polytechnique/Sorbonne Universit\'e, Universit\'e Paris Sud, Observatoire de Paris, Paris, France}
\author{O. Le~Contel}
\affiliation{Laboratoire de Physique des Plasmas,CNRS/Ecole Polytechnique/Sorbonne Universit\'e, Universit\'e Paris Sud, Observatoire de Paris, Paris, France}
\author{Y. Khotyaintsev}
\affiliation{Swedish Institute of Space Physics, Uppsala, Sweden}
\author{A. Vaivads}
\affiliation{Swedish Institute of Space Physics, Uppsala, Sweden}
\author{H. S. Fu}
\affiliation{School of Space and Environment, Beihang University, Beijing, China}
\author{F. Catapano}
\affiliation{Dipartimento di Fisica, Universit\`a della Calabria, Rende, Italy}
\affiliation{Laboratoire de Physique des Plasmas,CNRS/Ecole Polytechnique/Sorbonne Universit\'e, Universit\'e Paris Sud, Observatoire de Paris, Paris, France}
\author{H. Breuillard}
\affiliation{Laboratoire de Physique des Plasmas,CNRS/Ecole Polytechnique/Sorbonne Universit\'e, Universit\'e Paris Sud, Observatoire de Paris, Paris, France}
\affiliation{Laboratoire de Physique et Chimie de l'Environnement et de l'Espace, CNRS-Universit\'e d'Orl\'eans, France}
\author{N. Ahmadi}
\affiliation{Laboratory of Atmospheric and Space Physics, University of Colorado Boulder, Boulder, Colorado, USA}
\author{P.-A. Lindqvist}
\affiliation{KTH Royal Institute of Technology, Stockholm, Sweden}
%
%
\author{R. E. Ergun}
\affiliation{Laboratory of Atmospheric and Space Physics, University of Colorado Boulder, Boulder, Colorado, USA}
\author{R. B. Torbert}
\affiliation{Space Science Center, University of New Hampshire, Durham, New Hampshire, USA}
\author{B. L. Giles}
\affiliation{NASA Goddard Space Flight Center, Greenbelt, Maryland, USA}
\author{C. T. Russell}
\affiliation{Department of Earth and Space Sciences, University of California, Los Angeles, California, USA}

\author{R. Nakamura}
\affiliation{Space Research Institute, Austrian Academy of Sciences, Graz, Austria}
\author{S. Fuselier}
\affiliation{Southwest Research Institute, San Antonio, Texas, USA}
\affiliation{University of Texas at San Antonio, San Antonio, Texas, USA}
\author{B. H. Mauk}
\affiliation{The Johns Hopkins University Applied Physics Laboratory, Laurel, Maryland, USA}
\author{T. Moore}
\affiliation{NASA Goddard Space Flight Center, Greenbelt, Maryland, USA}
\author{J. L. Burch}
\affiliation{Southwest Research Institute, San Antonio, Texas, USA}

\date{\today}

\begin{abstract}
The Electron Diffusion Region (EDR) is the region where magnetic reconnection is initiated and electrons are energized. Because of experimental difficulties, the structure of the EDR is still poorly understood. A key question is whether the EDR has a homogeneous or patchy structure. Here we report Magnetospheric MultiScale (MMS) novel spacecraft observations providing evidence of inhomogeneous current densities and energy conversion over a few electron inertial lengths within an EDR at the terrestrial magnetopause, suggesting that the EDR can be rather structured. These inhomogenenities are revealed through multi-point measurements because the spacecraft separation is comparable to a few electron inertial lengths, allowing the entire MMS tetrahedron to be within the EDR most of the time. These observations are consistent with recent high-resolution and low-noise kinetic simulations.
\end{abstract}

\pacs{}

\maketitle


\section{Introduction} 

Magnetic reconnection is a fundamental energy conversion process occurring in space and laboratory plasmas  \cite{PriestForbes00, Vaivads06}. Reconnection occurs in thin current sheets leading to the reconfiguration of magnetic field topology and to conversion of magnetic energy into acceleration and heating of particles. Today, reconnection is recognized to play a key role in the Earth-solar environment, from the solar wind \cite{Phan06},  to magnetosheath \cite{Retino07, Phan18}, at the Earth's magnetopause \cite{Paschmann79, Vaivads04, Burch16} and in the magnetotail \cite{Oieroset01}. Reconnection is initiated in the Electron Diffusion Region (EDR), where electrons decouple from the magnetic field and are energized by electric fields \cite{Pritchett08}. Understanding the structure of the EDR is a key problem in reconnection physics which is still not solved. 

Pioneering spacecraft observations have provided partial evidence of the EDR \cite{Mozer03, Mozer05} in the Earth's subsolar magnetopause by showing theoretically predicted accelerated electrons, magnetic field-aligned currents and electric field on scales of electron skin depth. However, these observations lack time resolution for particle measurements.
Particle-in-Cell simulations of magnetopause reconnection have provided predictions of EDR signatures for the asymmetric case. These predictions include a peak of current density $\mathbf{J}$ \cite{Shay16}, non negligible electron agyrotropy $\sqrt{Q}$ \cite{Q, Swisdak16}, enhancements of parallel electron temperature, enhanced energy conversion $\mathbf{E}' \cdot \mathbf{J} \neq 0$ where $\mathbf{E}' =  \mathbf{E} + \mathbf{v}_e \times \mathbf{B}$,    \cite{Zenitani11}, non-negligible parallel (to the magnetic field) electric field \cite{Pritchett08} and meandering trajectories of electrons resulting in crescent-shaped distribution functions \cite{Hesse14, Bessho16, Lapenta17}. Another EDR evidence consists in the evolution of low energy field-aligned electron beams that are streaming towards the X-line in the IDR and that become oblique once they enter the EDR as they become demagnetized \cite{Egedal18}. 
Recent Magnetospheric MultiScale (MMS) mission measurements \cite{Burch16a}  have provided, for the first time, detailed evidence of the EDR at the magnetopause \cite{Burch16}. To date, several EDR encounters at the subsolar magnetopause have been reported \cite[and references therein]{Webster18} showing strong current densities of the order of $1000 \ nA/m^2$, electron agyrotropy $\sqrt{Q}$ up to $\sim 0.1$, parallel electron heating with $T_{e,||}/T_{e,\perp}$ up to $ \sim 4$, minima of $|\mathbf{B}| \sim 5 \ nT$, energy conversion $\mathbf{E}' \cdot \mathbf{J} \sim 10 \ nW /m^3$. Crescent-shape electron distribution functions are observed in most of cases and they are found on the magnetospheric side of the boundary \cite{Burch16}, in the electron outflow \cite{Norgren16} and in the magnetosheath inflow region \cite{Chen17}. \\
Until now, it is not fully understood whether the EDR has a preferred homogeneous or inhomogeneous structure at electron scales and below. EDR is identified as the site of strong vorticity \cite{Matthaeus82}. Also, current filamentation at electron scale can provide a source of anomalous resistivity leading to the violation of the frozen-in condition \cite{Che11}. Recent MMS observations of an EDR \cite{Burch16} have been compared to two-dimensional PIC simulations \cite{Shay16} and interpreted in terms of a laminar region. Yet, these simulations are two-dimensional, have limited spatial resolution and substantial averaging is performed in order to reduce noise. On the other hand, three-dimensional PIC simulations \cite{Daughton11, Price16, Price17}, two-dimensional PIC simulations with high spatial resolution \cite{JaraAlmonte14} or with low computational noise \cite{Swisdak18} indicate that the EDR can be rather inhomogeneous in electric fields, electron flows, current densities and energy conversion, with the formation of structures at electron-scale. Turbulent fluctuations, high vorticity and patchy energy conversion have been observed in the ion diffusion region \cite{Eastwood09,Fu17,Graham17,Ergun17} as well as in the outflow region \cite{Osman15, Phan16}. Recent observations \cite{Burch18} have shown that the presence of standing waves in the EDR leads to oscillatory energy conversion in the EDR. However, detailed observations supporting the structuring of the EDR are still lacking. \\
In this paper, we show MMS observations of an EDR encounter at the subsolar magnetopause when the four MMS probes were located at the smallest inter-spacecraft separation of  $\sim 6 \ km$, which is comparable to a few electron inertial length,  ($d_e\sim 2 \ km$). By comparing measurements of current, electric field,  energy conversion and electron distribution functions among the four spacecraft, we show that the EDR is structured at electron scales. A strong electron flow in the direction normal to the current sheet ($N$) leads to a non-zero energy conversion in that direction ($E'_N  J_N$) which is inhomogeneous and comparable to the contribution of the energy conversion $E'_M J_M$ where $M$ is the direction parallel to the current in the current sheet. These inhomogeneities can be revealed through multi-point measurements only when the spacecraft separation is comparable to a few  electron inertial lengths, since the entire MMS tetrahedron is within the EDR. In these observations, the separation is $\sim 3$ electron inertial lengths.


\section{Observations}

\subsection{Electron Diffusion Region signatures}

MMS spacecraft \cite{Burch16a} encountered the EDR on January, 27$^{th}$ 2017, during a magnetopause crossing taking place between 12:05:41.9 and 12:05:44.0 UTC. At that time, the MMS constellation was located in the subsolar magnetopause region, at $(9.3, \  -1.2, \   2.1) \ R_E$ in Geocentric Solar Ecliptic (GSE) coordinates. The mean spacecraft separation was $\sim 6 \ km$, which is the smallest possible for MMS. Figure 1 shows a $5$ minute interval that includes the EDR crossing marked by the yellow shaded region. Fig.1a shows the magnetic field components measured by the WIND spacecraft \cite{Acuna95} in the solar wind, which have been shifted by 47 minutes to take into account propagation to the magnetopause. Fig.1b-d show the MMS1 measurements of the magnetic field components, ion density and ion velocity components in the GSE coordinate system. Throughout the paper, the burst mode data are used: the magnetic field data from the FluxGate Magnetometer (FGM) at $128$ samples/s \cite{Russell2016}, 3D electric field data from the axial \cite{Ergun16} and spin-plane \cite{Lindqvist16} probes at $8192$ samples/s and particles data from the Fast Plasma Instrument (FPI) with $30 \ ms$ for electrons and $150 \ ms$ for ions \cite{Pollock16}. Throughout the paper, current densities are computed using single spacecraft data at the electron resolution ($30 \ ms$), $\mathbf{J} = e \ n_e (\mathbf{v}_i - \mathbf{v}_e)$. MMS stays mostly in the magnetospheric boundary layer,  which corresponds to $B_z > 0$ (Fig. 1b) and to the typical value of the density $\sim 10 \ cm^{-3}$ (Fig.1c) \cite{Eastman79}. Between 12:05:41.2 and 12:05:43.2, $B_z$ becomes negative. Fig.1a shows that the magnetic field in the magnetosheath adjacent to the magnetopause was stable and directed southward, supporting the fact that when $B_z < 0$ MMS is on the magnetosheath side of the magnetopause boundary.
An ion and electron $v_z$ jet reversal are observed at the second $B_z$ reversal, at 12:05:43.20 (Fig.1d and Fig.1f).
The ion velocity in the $z$ direction changes from a value of $+200 \ km/s$ (12:05:41.0) to $-150 \ km/s$ (12:05:48.0). The jet reversal is observed also in the electron velocity and $v_{e,z}$ changes from $\sim +250 \ km/s$ to $\sim -450 \ km/s$ (the local ion Alfv\`en speed is $\sim 100 \ km/s$).   The high speed ion and electron flows, the corresponding ion and electron flow reversals as well as the $B_z$ reversal and the low $|\mathbf{B}| \sim 3 \ nT$ indicate that the spacecraft is in the vicinity of the reconnection region at  12:05:41.9 - 12:05:48.0 (yellow shaded region in Fig.1a-1d).

\begin{figure}
	\centering
	\includegraphics[width=1\columnwidth]{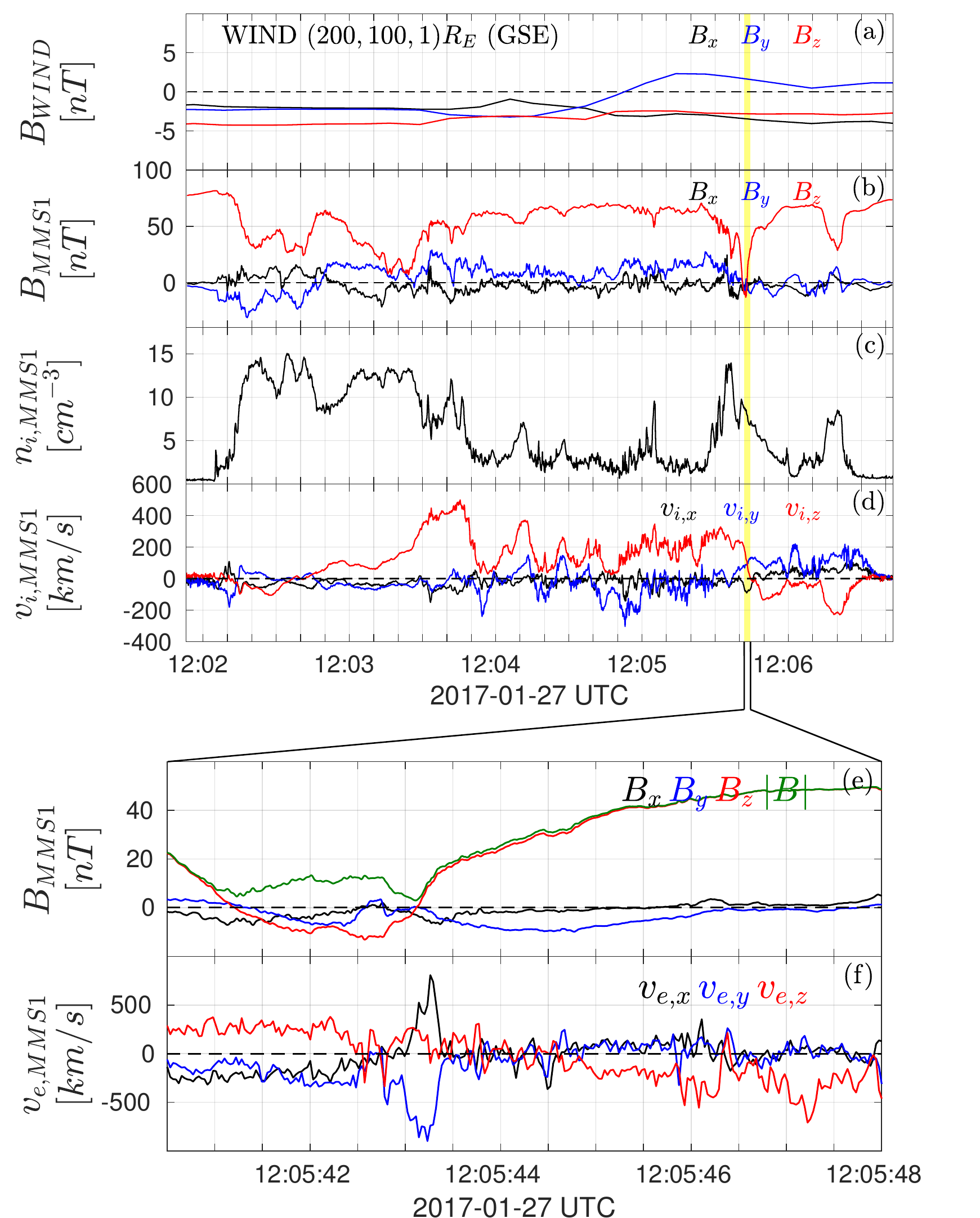}
	\caption{(a) Magnetic field components as measured by WIND and propagated to the magnetopause; (b) MMS1 magnetic field components; (c) MMS1 ion density and (d) MMS1 ion velocity components; (e) Zoom-in of the MMS1 magnetic field components and strength; (f) Zoom-in of the electron velocity components. Data are shown in GSE. The yellow shaded region in panel(a)--(d) indicates the EDR crossing. \label{fig1}}
\end{figure}


The approximate  trajectory of the spacecraft through the reconnection region is shown in Fig.2. From Fig.2 onwards, data are shown in the local current sheet coordinate system, LMN. The LM plane represents the current sheet plane, where M is the direction parallel to the current, and N is perpendicular to the current sheet. In the GSE coordinates, L = (-0.039,  -0.252,    0.967), which is close to the south-north direction, M = (-0.301,   -0.921,  -0.252), which is approximately the east-west direction, and N = (0.954,   -0.300,   -0.040), which is approximately parallel to the Earth-Sun direction. The local reference frame LMN is obtained applying the Minimum Variance Analysis on the $\mathbf{B}$ data in the interval 12:05:41.9 - 12:05:46.9. The eigenvalue ratios for MMS4 are $\lambda_L/\lambda_M \sim 40$ and $\lambda_M/\lambda_N \sim 10$. The single spacecraft LMN systems are then averaged over the four spacecraft. An additional rotation of $17^{\circ}$ around the N direction is added in order to guarantee the consistency of $B_M$ and $J_L$ measurements within the diffusion region with the Hall pattern.

In the interval shown in Fig.2 (12:05:41.9 - 12:05:44.0), ions are not magnetized (see Fig.3d) and $B_M$ (Fig.2b) corresponds to the out-of-plane Hall field with a distorted quadrupolar pattern, as expected for asymmetric reconnection with a weak guide field \cite{Chen17}, with $B_M > 0$ ($B_M < 0$) on the magnetosheath side of the boundary, northern (southern) of the reconnection site. These observations indicate that the spacecraft is located in the ion diffusion region. The guide field is estimated to be less than $10 \%$ of $|\mathbf{B}|$ according to the averaged value of $B_M$ among the spacecraft in the center of the current sheet ($B_L$ inversion).

In interval AB (12:05:41.900 - 12:05:42.456,  Fig.2), all four probes observe roughly constant values of $B_L<0 $ yet showing differences of several $nT$ despite the small inter-spacecraft separation, indicating  that the current sheet is thin. A large parallel current ($J_L < 0$ in Fig.2d) and Hall magnetic field $B_M > 0$ (Fig.2b) indicate that MMS is close to the current sheet on the magnetosheath side of the boundary, north of the reconnection site. The probes are rather close to the center of the current sheet, as indicated by the large $J_M \sim 500 \ nA/m^2$ and small $B_L$. According to the $B_L$ difference among the probes, MMS3 is the closest to the center of the current sheet (see the tetrahedron close to location A in Fig.2g) while MMS4 and MMS1 are further away.
In this interval, the trajectory of MMS is tangential to the magnetopause, therefore differences among the spacecraft observations have to be considered as spatial.

In interval BC (12:05:42.456 - 12:05:42.830), the peaks of $J_L > 0$ indicate that MMS moves closer to the magnetosheath separatrix. MMS1 and MMS4 make a brief excursion in the inflow region around 12:05:42.6, where the $B_L$ gradient is smaller and all probes except MMS3 observe a minimum in $J_M$ and $B_M \sim 0$. At the same time MMS3, which is closer to the center of the current sheet, observes $B_M \sim 4 \ nT$ and large $J_M$. Accordingly, the location of the four spacecraft at this time is shown in Fig.2g with the projection of the tetrahedron in the plane LN between the letters B and C indicating the corresponding time interval. 
After that, MMS1 and MMS4 cross again the magnetospheric separatrix and the constellation comes back in the Hall region where $B_M \sim 5 \ nT$ for all the spacecraft (at 12:05:42.830).

In interval CD (12:05:42.83 - 12:05:43.65), MMS crosses the current sheet north of the reconnection site ($B_N<0$). By applying the timing method \cite{Paschmann98} to this current sheet crossing, we estimate the normal velocity of the current sheet  to be about $\sim 35 \ km/s$ and the normal direction to be $\mathbf{n} = (0.95, \ 0.25, \ 0.08)$ (GSE). The normal direction, estimated by timing is  in good agreement with the  normal found with the MVA method. According to the current sheet speed, MMS crosses an electron scale current sheet with a thickness of $\sim 30  \ km \sim 15 \ d_e$. The current sheet corresponds to a strong value of $J_M > 1000 \ nA/m^2$. The strong decrease in $B_N$ in the CD interval corresponds to the reconnected magnetic field.
The curvature radius of the magnetic field lines $R_c = \mathbf{b} \cdot \nabla \mathbf{b}$ (where $\mathbf{b} = \mathbf{B}/|\mathbf{B}|$, Fig.2f) decreases as well reaching its minimum of less than $10 \ km \sim 5 \ d_e$ at the $|\mathbf{B}|$ minimum ($\sim 3 \ nT$). This indicates that the spacecraft is located close to the center of reconnection site at this time.
Furthermore, The FOTE method \cite{Fu15} applied to this event (not shown) indicates that the minimum distance between the spacecraft and the null point is $\sim 12 \ km \sim 6 \ d_e$.

After the current sheet crossing (CD interval), MMS moves tangentially along the southern magnetospheric separatrix region observing a southward ion and electron jet $v_{i,L}, v_{e,L} <0$ (corresponding to $v_{i,z}$ and $v_{e,z}$ in Fig.1d and Fig.1f).

\begin{figure}
	\centering
	\includegraphics[width=1\columnwidth]{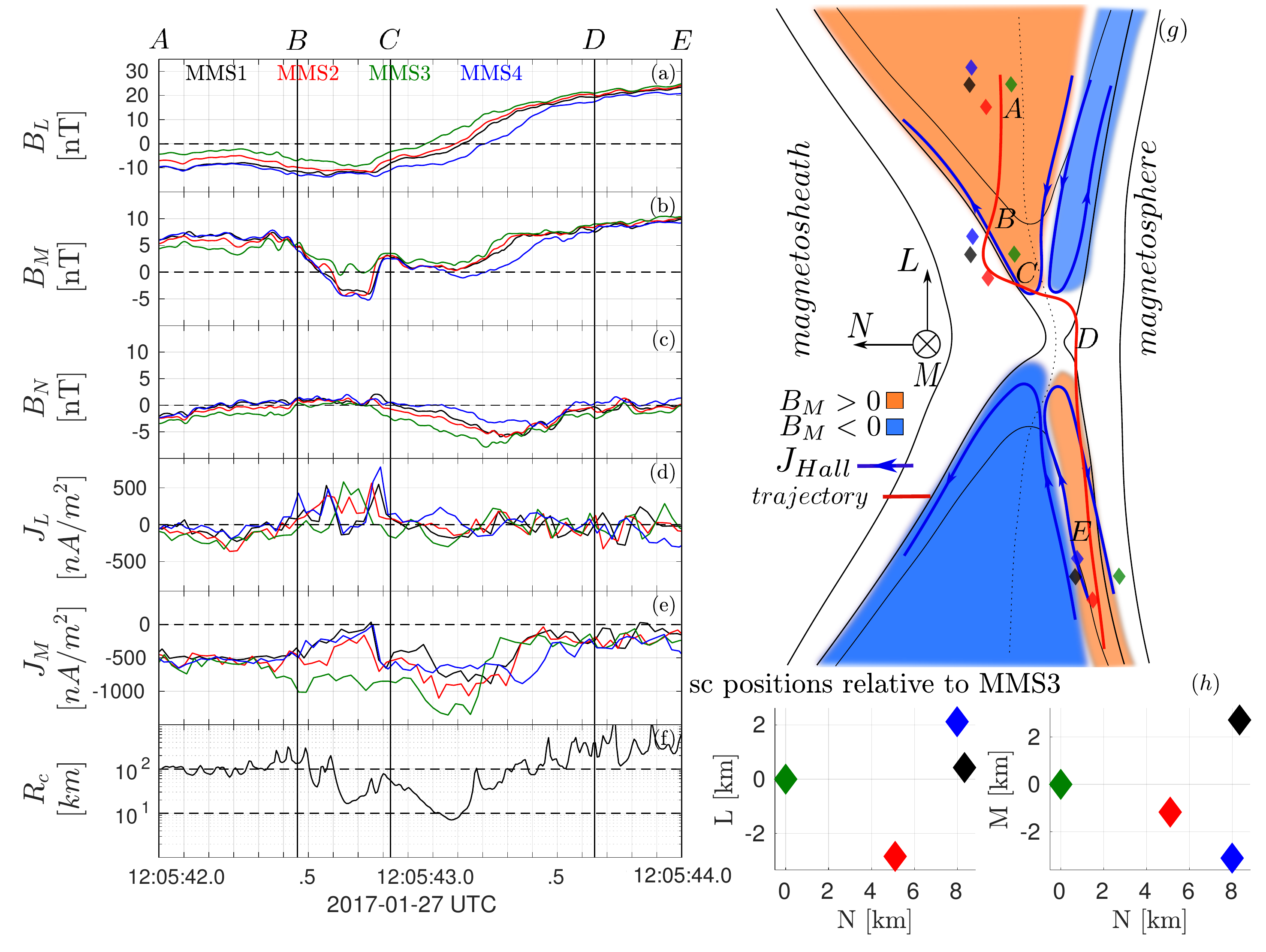}
	\caption{Four spacecraft measurements of (a) $B_L$; (b) $B_M$; (c) $B_N$; (d) $J_L$; (e) $J_M$. (f) Curvature radius of the magnetic field lines; (g) cartoon of the encounter. The red line represent the trajectory of the barycenter of MMS constellation. Since the velocity of the magnetopause is much larger than the spacecraft velocities, the MMS path shown is produced entirely by the motion of the magnetopause in the LN plane. The three tetrahedra represent MMS location at different times along the trajectory; (h) Projection of the MMS tetrahedron in the LN and in the MN plane. The tetrahedron quality factor is $0.84$.        
		\label{fig2}}
\end{figure}


The schematic trajectory of MMS (Fig.2g) indicates that the spacecraft crossed the magnetopause close to the reconnection site. Figure 3 shows further evidence of MMS crossing the EDR. During the magnetopause crossing identified by the $B_L$ reversal (Fig.3a), a large enhancement of the electron velocity shifted toward the magnetosphere is observed in $M$ and $N$ components, reaching $600 \ km/s$ and $1000 \ km/s$ respectively (Fig.3b). These peaks are not observed in the ion velocity. Therefore, the current densities presented in Fig.3c are carried by electrons and they peak between 12:05:43.200 and 12:05:43.350 reaching $\sim 1000 \ nA/m^{2}$ in $J_M$ and $J_N$. These values of $J_M$ are expected for a current sheet at the electron scales and similar values are reported in other EDR observations \cite{Burch16, Webster18}. A further confirmation of the EDR encounter is given by the demagnetization of electrons (Fig.3d), which are decoupled from the magnetic field ($\mathbf{E} \neq - \mathbf{v}_e \times \mathbf{B} $) between 12:05:43.150 and 12:05:43.350.  
Consistently with the trajectory in Fig.2, a positive $v_{e,L} \sim 400 \ km/s$ is observed between 12:05:42.900 and 12:05:43.250 and $v_{e,L} \ll v_{A,e} \sim 4000 \ km/s$, the electron Alfv\'en speed. This indicates that MMS is crossing the inner EDR, where the electron jet has not developed yet \cite{Karimabadi07}.
Agyrotropy $\sqrt{Q}$  (Fig.3e) \cite{Q, Swisdak16} exhibits an enhancement in correspondence of the $B_L$ reversal. The agyrotropy parameter $\sqrt{Q}$ can have non negligible values also far from the EDR, specifically along the magnetospheric separatrix \cite{Lapenta17}[e.g. Fig.3], \cite{Shay16}.  Yet in the present case, the agyrotropy increase is observed by all four MMS probes between 12:05:42.6 and 12:05:43.5 and for the majority of this interval (12:05:42.6 - 12:05:43.2) MMS is in the magnetosheath ($B_L < 0$). The electron temperature increase is shifted towards the magnetosphere and mainly seen in the direction parallel to the magnetic field \cite{Shay16, Egedal11} ($\Delta T_{e,||} \sim 50 \ eV$ and $\Delta T_{e,\perp} \sim 25 \ eV$  through the crossing) while at the $|\mathbf{B}|$ minimum $T_{e,||} \sim T_{e,\perp}$. The same behavior is shown also by the electron Pitch Angle Distribution (PAD) (Fig.3g). Furthermore, between 12:05:42.760 and 12:05:42.980 a low energy electron population parallel to $\mathbf{B}$ propagates toward the $|\mathbf{B}|$ minimum. At the $|\mathbf{B}|$ minimum (12:05:42.980 - 12:05:43.150) this beam is no longer observed and the PAD looks isotropic while the distribution functions exhibit oblique beams (to the magnetic field). This signature has been recently identified as the indication of electron demagnetization \cite{Egedal18}. In addition, the strong fluctuations in the electric field data observed in correspondence of the $\mathbf{|B|}$ minimum (Fig.4e-f) suggest that high frequency waves may be present.
All  these EDR encounter signatures are shown using MMS1 data and they were observed overall by all probes, albeit with some differences which are significant and will be discussed below.

\begin{figure}
	\centering
	\includegraphics[width=1\columnwidth]{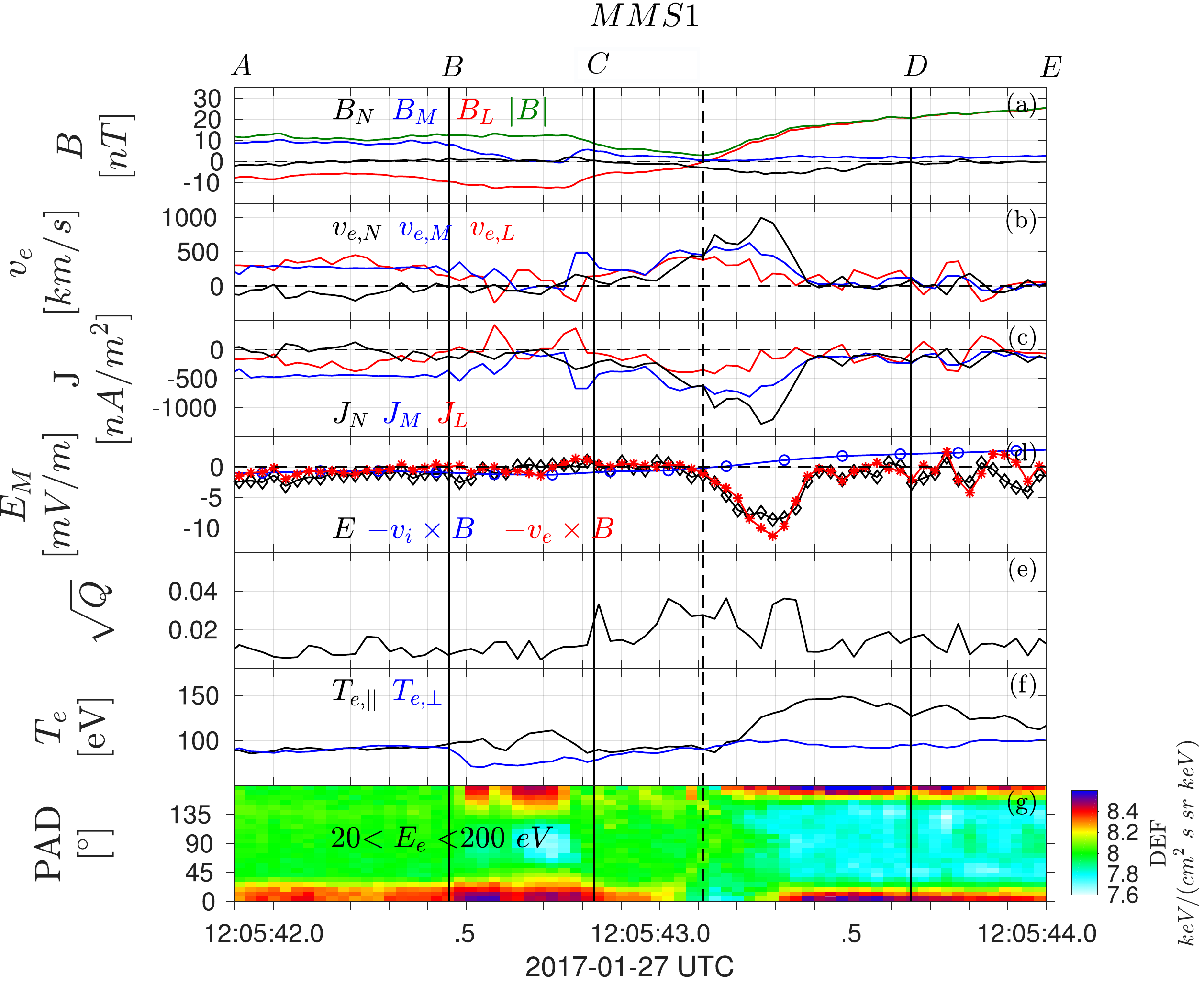}
	\caption{   (a) Magnetic field components and strength; (b) electron velocity components; (c) current density components; (d) M component of electric field ($30 \ ms$ resolution), $(\mathbf{v}_e \times \mathbf{B})_M$ ($30 \ ms$ resolution), $(\mathbf{v}_i \times \mathbf{B})_M$ ($150 \ ms$ resolution); (e) agyrotropy parameter $\sqrt{Q}$; (f) parallel and perpendicular electron temperature; (g) electron pitch angle distribution in the energy range $[20, 200] \ eV$. The black vertical dashed-line indicates the time of the $|\mathbf{B}|$ minimum. Data from MMS1.  \label{fig3}}
\end{figure}


\subsection{Electron-scale structuring of the EDR}

Figure 4 show the four-spacecraft analysis of the EDR encounter.
Fig.4a and Fig.4b show respectively $B_L$ measured by each spacecraft and the shifted $B_L$ obtained via the timing method \cite{Paschmann98}. The time lag between $B_L$ components measured by MMS1 and MMS2-3-4 respectively are $\Delta t = (\Delta t_{12},\Delta t_{13}, \Delta t_{14})  = (0.024 \ s, 0.114 \ s, -0.113 \ s)$. In order to facilitate the comparison among observations by different spacecraft, the same shift is applied to Fig.4c-4i. We note that all the probes observe a large $J_M$ consistent with the current sheet crossing. However, while $J_M$ reaches $1200 \ nA/m^2$ for MMS3, its value is lower ($\sim 800 \ nA/m^2$) for the other probes. The difference in the current density observations by different MMS probes is larger than the FPI measurement error, which is $\sim 20 \%$ \cite{Pollock16}. Therefore, the current densities in the EDR are not homogeneous on the scale of a few $d_e$, which corresponds to the spacecraft separation. To summarize, we may say that at large ion scales the current densities are homogeneous, while by looking  at the electron scale we are able to observe fine structures that may be due to the filamentation of the current sheet (see Fig.4k, upper right frame).
The electric field $E_M$ (Fig.4e) and $E_N$ (Fig.4f) maintain the same sign during the EDR crossing. $E_M$ and $E_N$ are comparable and they both reach $10 \ mV/m$. This differs from what is expected in the case of laminar and steady two-dimensional reconnection, where close to the reconnection site $E_M$ represent the reconnection electric field and it is typically much smaller than the Hall field $E_N$. 
Fig.4d shows that a large peak of $J_N \sim -1000 \ nA/m^2$ is seen by all the spacecraft. Such a large $J_N<0$ corresponds to a large $v_{e,N}$ directed toward the magnetosheath. Note that this $J_N$ behavior is not typically observed close to the reconnection site in two-dimensional PIC simulations \cite{Pritchett08, Shay16} and observations \cite{Burch16}. Since the $v_{e,N}>0$ region is observed by all spacecraft, its minimal width has to be comparable to the spacecraft separation. In particular, in the LN plane, the minimal width of the $v_{e,N} > 0$ region is $4 \ km \sim 2 \ d_e$ in the L direction and of $8 \ km \sim 4 \ d_e$ in the N direction.

The strong $J_N$ deeply affects the energy conversion pattern since $E'_N J_N$ (Fig.4h) becomes comparable to $E'_M J_M$ (Fig.4g). 
If we consider the maximum error associated to each quantity (with $\delta E = 20 \% |E|$, $\delta B = 0.5 \ nT$ and an error of  $\sim 10 \%$ for density and velocity) we find that $E'_M J_M$ has a positive peak for MMS3 while for MMS4 $E'_M J_M$ shows a bipolar signature that is beyond the errors (Fig.4g). In Fig.4g-i only data from MMS3 and MMS4 are shown since they exhibit the clearest differences between spacecraft. All four probes quantities and associated errors are shown in the Supplementary Material. The energy conversion errors are comparable to the measured quantities for all the spacecraft. However, on MMS4 errors are smaller so that we obtain an unambiguous value for the total $\mathbf{E}' \cdot \mathbf{J} \sim E'_M J_M + E'_N J_N$  ($E'_L J_L \ll E'_M J_M, E'_N J_N$). In particular, on MMS4 $\mathbf{E}' \cdot \mathbf{J}<0$ (Fig.4i), showing negative energy transfer between fields and particles. This indicates that energy is locally converted from the particles to the field, the opposite of the standard behavior during reconnection. This is sketched in the bottom right panel of Fig.4k.
Since MMS4 is the only spacecraft that provides a value of the energy conversion $\mathbf{E}' \cdot \mathbf{J}$ beyond the errors, we have also computed the electric field using Ohm's law

\begin{equation}
\mathbf{E}'_{FPI} =  - \frac{\mathbf{\nabla} \overline{P}_e}{n e} + \frac{m_e}{e} \mathbf{v}_e \cdot \mathbf{\nabla} \mathbf{v}_e + \frac{m_e}{e} \frac{\partial}{\partial t} \mathbf{v}_e
\end{equation}

Here, $\overline{P}_e$ is the electron pressure tensor and the subscript $FPI$ indicates that $\mathbf{E}'_{FPI}$  is obtained by using measurements from FPI instrument only. $\mathbf{\nabla} \overline{P}_e$ is calculated using four spacecraft measurements and the full pressure tensor \cite{Paschmann98} so it is an average over the spacecraft tetrahedron. Note that the errors on particles data provided by FPI \cite{Pollock16} are smaller than the electric field errors.
We found that, since the contribution of the inertia term is negligible (not shown), a good proxy for the electric field is $\mathbf{E}'_{FPI} = - \mathbf{\nabla} \overline{P}_e / n e $. The quantities $E'_{FPI,N} J_N$ (Fig.4h) and $E'_{FPI,M} J_M$ (Fig.4g), exhibit bipolar signatures, as the total energy conversion $\mathbf{E}'_{FPI} \cdot \mathbf{J}$ (Fig.4j). Yet, it should be noted that $\mathbf{E}'_{FPI}$ is a four-spacecraft measurement averaged over the tetrahedron and one should be careful when comparing it to single spacecraft observations especially if, as in this case, significant differences are seen among probes' observations. For consistency, $\mathbf{J}$ is the current density which is also averaged over the tetrahedron in this case. After a careful evaluation of all error sources, we conclude that the discrepancy between the \textit{punctual} (as given by MMS4) and the \textit{averaged} energy conversion (given by $\mathbf{E}'_{FPI} \cdot \mathbf{J}$) is not an instrumental effect and indicates that energy conversion is not homogeneous over the tetrahedron and that energy conversion is patchy over scales of the order of few $d_e$.

The evolution of the electron distribution functions (DFs) measured by MMS4 in the EDR is shown in Fig.4l-t. The projection of the electron DFs are made in the three perpendicular planes $(\mathbf{v}_{\perp,1}, \mathbf{v}_{\perp,2})$, $(\mathbf{v}_{||}, \mathbf{v}_{\perp,2})$ and $(\mathbf{v}_{||}, \mathbf{v}_{\perp,1})$ where $\mathbf{v}_{\perp,1}= \mathbf{v} \times \mathbf{b}$, $\mathbf{v}_{\perp,2} = \mathbf{b} \times (\mathbf{v} \times \mathbf{b})$ and $\mathbf{v}_{||}=\mathbf{b}$ ($\mathbf{v} = \mathbf{v}_e/|\mathbf{v}_e|$ and  $\mathbf{b}=\mathbf{B}/|\mathbf{B}|$) and at the three times indicated by the vertical black lines in Fig.4b-4i. The times are shifted according to the delays among spacecraft obtained with the timing method. These times correspond to regions where $E'_M J_M$ is positive (DFs indicated with $\alpha$, Fig.4l-n), negative (DFs indicated with $\beta$, Fig.4r-t) and in the transition from positive to negative (DFs indicated with $\gamma$, Fig.4o-q). Similar DFs are observed by all spacecraft and they last for more than one FPI measurement (with $30 \ ms$ resolution). The $\alpha$ DFs (Fig.4l-n) have a rather complicated shape with several oblique beams.
This pattern is observed around the magnetic field minimum, from 12:05:43.179 to 12:05:43.269 for MMS4. When $E'_M J_M$ changes sign, the DFs change shape (Fig.4o-q) and clearly become crescent-shaped distributions in the $(\mathbf{v}_{\perp,1}, \mathbf{v}_{\perp,2})$ plane when $E'_M J_M<0$ (Fig.4r). The DFs observed during this EDR encounter are rather complex. They are not always crescent-like and they appear to be related to $E'_M J_M$. Further analysis and comparisons with simulations are needed to fully understand the dynamics of electrons in such a complex EDR.    

\section{Discussion and Conclusions}

We have presented observations of an Electron Diffusion Region (EDR) encountered at the magnetopause by the MMS spacecraft with the very low inter-spacecraft separation of $\sim  3$ electron inertial length.
During this electron-scale current sheet crossing the four MMS spacecraft observe typical EDR signatures \cite{Webster18} suggesting that MMS crossed the magnetopause in close proximity to a X-line. These signatures include a large current density mainly carried by electrons (Fig.3b-3c), a peak of electron agyrotropy (Fig.3e), demagnetization of ions and electrons (Fig.3d-3g), increased electron temperature anisotropy with $T_{e,||} > T_{e, \perp}$ (Fig.3f), crescent-shaped electron distribution functions (Fig.4o-4r). Furthermore, we observe that the electron jet has not fully developed ($v_{A,i} < v_{e, L} < v_{A,e}$) indicating that MMS is within the inner EDR \cite{Karimabadi07}.

Another observed inner EDR signature is the fact that low energy field-aligned electron beams directed towards the X-line become oblique in close proximity to the center of the EDR (Fig.3g). This behaviour indicates electron demagnetization. Indeed, 2D kinetic simulations \cite{Egedal18} showed that the transition from the field-aligned distribution to the one with oblique beams takes place where the magnetic field is sharply changing direction and has the smallest magnitude, leading to the electron decoupling from the magnetic field.

In the presented event, all four MMS probes observed the EDR signatures. The multi-spacecraft analysis of the EDR revealed that the current density $J_M$ is spatially inhomogeneous at electron scales (Fig.4c-4k). Previously reported EDR encounters \cite{Burch16, Chen17} do not point out differences among spacecraft in the current density because either the inter-spacecraft separation was not small enough to have all the spacecraft within the EDR and to resolve the electron scale inhomogeneities, either the EDR becomes structured at electron scale only under particular conditions (e.g. depending on the guide field value and on the inflow condition). Indeed, similar inhomogeneities have been seen in high-resolution PIC simulations \cite{JaraAlmonte14} where the current density is found to be structured at electron scale and below.

Strikingly, in the center of the reconnection site, the current density in the direction normal to the current sheet, $J_N$, is observed to have almost the same magnitude as the out-of-plane $J_M$ current density (Fig.4c-4d). In addition, electrons are observed to move from the magnetosphere to the magnetosheath side of the magnetopause, corresponding to $J_N<0$. This behaviour of electrons differs from  the  typical observations close to the reconnection site \cite{Burch16} as well as predictions by 2D PIC simulations as e.g.   \cite{Pritchett08, Shay16}. However, our observations are consistent with recent PIC simulations with low numerical noise \cite{Swisdak18, Egedal18} in which electrons move downstream along the magnetospheric separatrix performing oscillations of decaying amplitude in the $N$ direction. The velocity oscillations observed in simulations  \cite{Swisdak18, Egedal18} are composed by alternating regions, or channels, of positive and negative $v_{e,N}$. In the EDR encounter presented here, such oscillations are not observed (Fig.4d), which might indicate that all the spacecraft were measuring the same channel with $v_{e,N}>0$. Accordingly, we infer that the channel’s width has to be comparable to or larger than the inter-spacecraft separation of $\sim 3 \ d_e$.

Another characteristic of the presented EDR is the similarity in magnitude of the electric field $E_M$ and $E_N$ components. This has been identified as one of the signatures of inhomogeneous current layer ‘’disrupted’’ by turbulence in three-dimensional simulations \cite{Price16}. Accordingly, our observations support the picture of the EDR as the site of strong spatial gradients and inhomogeneities.

The energy conversion  $\mathbf{E}' \cdot \mathbf{J}$ (Fig.4i) is highly affected by the  $J_N \sim J_M$ and $E_M \sim E_N$ behavior since the two terms $E'_M J_M$ and $E'_N J_N$ become comparable  (Fig.4g-4h). In other EDR encounters by MMS \cite{Burch18, Burch16}, $\mathbf{E}' \cdot \mathbf{J} \sim E'_M J_M$ since $J_N$ is usually negligible in comparison to $J_M$. For the EDR presented here, the multi-spacecraft analysis revealed that energy conversion $\mathbf{E}' \cdot \mathbf{J}$ is spatially inhomogeneous at electron scales. We have also shown that the quantitative evaluation of energy conversion, is affected by the experimental errors (Fig.4g-i). However, the comparison of the single spacecraft measurements from different spacecraft (Fig.4g-i) and the measurements averaged over the tetrahedron (Fig.4j) both support the qualitative picture in which $\mathbf{E}' \cdot \mathbf{J}$ is patchy and changing sign in the vicinity of the reconnection site. This implies that the EDR comprises of regions where energy is transferred from the field to the plasma and regions with the opposite energy transition, which is unexpected during reconnection. A negative energy conversion was also observed in the outer EDR \cite{Hwang17}.

Electron-scale variations of $\mathbf{E}' \cdot \mathbf{J}$ in the EDR have been recently observed \cite{Burch18}. However, in \cite{Burch18} these variations are oscillations of $\mathbf{E}' \cdot \mathbf{J}$ which are the consequence of the oscillatory electric field pattern that shows signatures of a standing wave. This differs from the $\mathbf{E}' \cdot \mathbf{J}$ behavior reported in our study where no such oscillatory behavior of the electric field is observed and the patchy energy conversion is consistent to spatial inhomogeneities due to electron scale structuring.

The origin of the patchy energy conversion appears to be connected to the large $v_{e,N} \sim v_{e,M}$ directed from the magnetosphere to magnetosheath that has never been observed before. 
Further observational cases as well as 3D PIC simulations with higher resolution and lower noise or full Vlasov simulations are required to understand which conditions may lead to the structuring of the EDR and how this patchy structure affect the electron energization. These observations can be an indication of what  might be observed in the EDR in the magnetotail, where highly detailed observation are available since the inter-spacecraft separation of MMS is of the order of $1 \ d_e$.

\begin{figure}
	\centering
	\includegraphics[width=1\columnwidth]{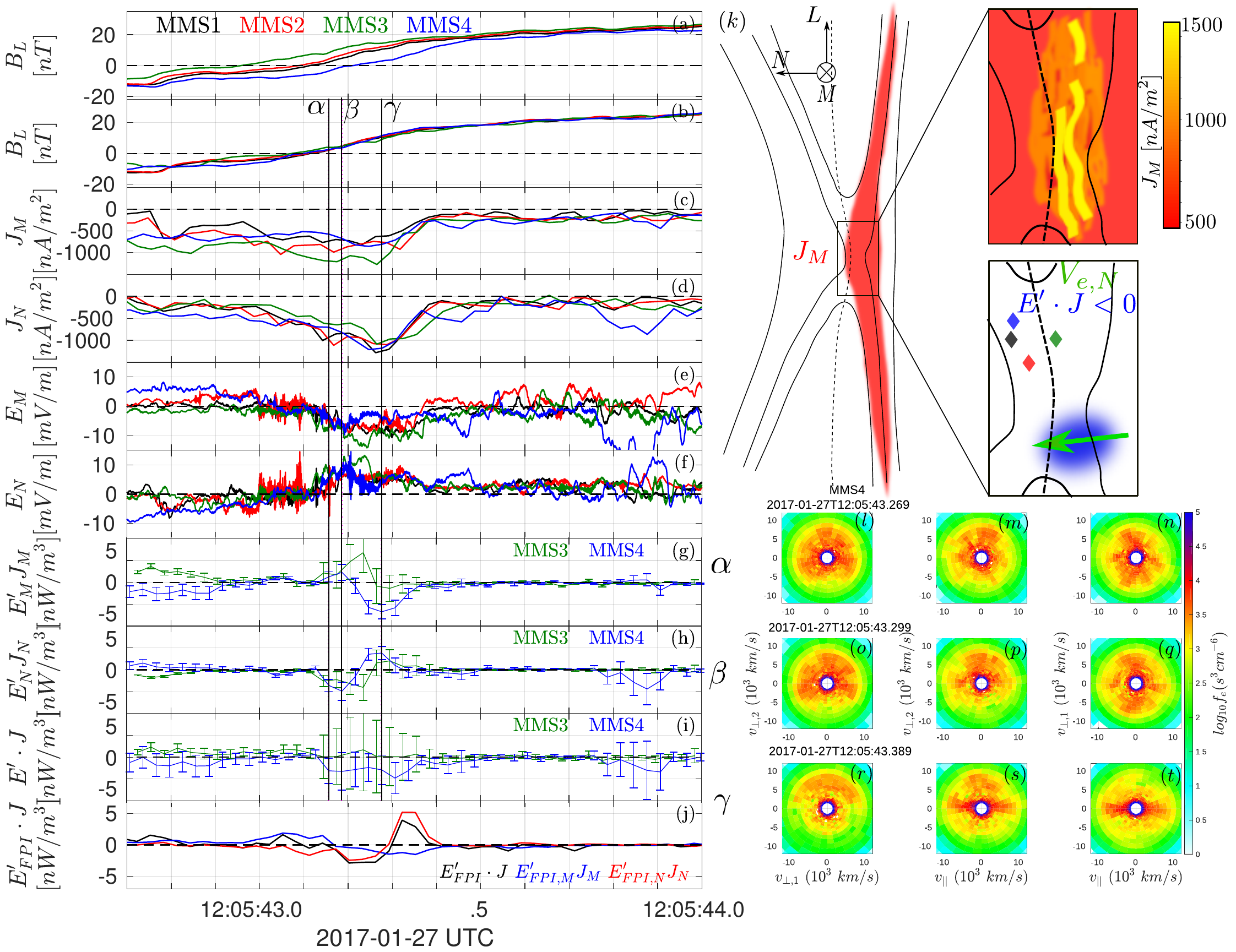}
	\caption{Four spacecraft (a) $B_L$; (b) Time-shifted $B_L$. (c) Time-shifted $J_M$; (d) Time-shifted $J_N$; (e) Time-shifted $E_M$ (8192 samples/s); (f) Time-shifted $E_N$ (8192 samples/s); (g) Time-shifted $E'_M J_M$; (h) Time-shifted $E'_N J_N$; (i) Time-shifted $\mathbf{E}' \cdot \mathbf{J}$; (j) $\mathbf{E}'_{FPI} \cdot \mathbf{J}$, $E'_{FPI,M} J_M$, $E'_{FPI,N} J_N$. The $\alpha$, $\beta$ and $\gamma$ lines correspond to the times of the  $\alpha$, $\beta$ and $\gamma$  distribution functions in panels (l)-(t) shifted accordingly to the timing method. (k) Cartoon of $J_M$ and of the energy conversion and $v_{e,N}$ flow. (l)-(t) Electron distributions by MMS4 projected on $(\mathbf{v}_{\perp,1}, \mathbf{v}_{\perp,2})$, $(\mathbf{v}_{||}, \mathbf{v}_{\perp,2})$ and $(\mathbf{v}_{||}, \mathbf{v}_{\perp,1})$ planes at three different times $t_{\alpha} =$ 12:05:43.269, $t_{\beta} =$ 12:05:43.299, $t_{\gamma} =$ 12:05:43.389.   
		\label{fig4}}
\end{figure}

\begin{acknowledgments}
	We thank the entire MMS team and instruments PIs for data access and support. MMS data are available at https://lasp.colorado.edu/mms/sdc/public.
\end{acknowledgments}


\begin{thebibliography}{48}%
	\makeatletter
	\providecommand \@ifxundefined [1]{%
		\@ifx{#1\undefined}
	}%
	\providecommand \@ifnum [1]{%
		\ifnum #1\expandafter \@firstoftwo
		\else \expandafter \@secondoftwo
		\fi
	}%
	\providecommand \@ifx [1]{%
		\ifx #1\expandafter \@firstoftwo
		\else \expandafter \@secondoftwo
		\fi
	}%
	\providecommand \natexlab [1]{#1}%
	\providecommand \enquote  [1]{``#1''}%
	\providecommand \bibnamefont  [1]{#1}%
	\providecommand \bibfnamefont [1]{#1}%
	\providecommand \citenamefont [1]{#1}%
	\providecommand \href@noop [0]{\@secondoftwo}%
	\providecommand \href [0]{\begingroup \@sanitize@url \@href}%
	\providecommand \@href[1]{\@@startlink{#1}\@@href}%
	\providecommand \@@href[1]{\endgroup#1\@@endlink}%
	\providecommand \@sanitize@url [0]{\catcode `\\12\catcode `\$12\catcode
		`\&12\catcode `\#12\catcode `\^12\catcode `\_12\catcode `\%12\relax}%
	\providecommand \@@startlink[1]{}%
	\providecommand \@@endlink[0]{}%
	\providecommand \url  [0]{\begingroup\@sanitize@url \@url }%
	\providecommand \@url [1]{\endgroup\@href {#1}{\urlprefix }}%
	\providecommand \urlprefix  [0]{URL }%
	\providecommand \Eprint [0]{\href }%
	\providecommand \doibase [0]{http://dx.doi.org/}%
	\providecommand \selectlanguage [0]{\@gobble}%
	\providecommand \bibinfo  [0]{\@secondoftwo}%
	\providecommand \bibfield  [0]{\@secondoftwo}%
	\providecommand \translation [1]{[#1]}%
	\providecommand \BibitemOpen [0]{}%
	\providecommand \bibitemStop [0]{}%
	\providecommand \bibitemNoStop [0]{.\EOS\space}%
	\providecommand \EOS [0]{\spacefactor3000\relax}%
	\providecommand \BibitemShut  [1]{\csname bibitem#1\endcsname}%
	\let\auto@bib@innerbib\@empty
	\bibitem [{\citenamefont {{Priest}}\ and\ \citenamefont
		{{Forbes}}(2000)}]{PriestForbes00}%
	\BibitemOpen
	\bibfield  {author} {\bibinfo {author} {\bibfnamefont {E.~R.}\ \bibnamefont
			{{Priest}}}\ and\ \bibinfo {author} {\bibfnamefont {T.}~\bibnamefont
			{{Forbes}}},\ }\href@noop {} {\emph {\bibinfo {title} {Magnetic reconnection.
				MHD theory and applications}}}\ (\bibinfo  {publisher} {Cambridge University
		Press},\ \bibinfo {address} {United Kingdom},\ \bibinfo {year}
	{2000})\BibitemShut {NoStop}%
	\bibitem [{\citenamefont {Vaivads}\ \emph {et~al.}(2006)\citenamefont
		{Vaivads}, \citenamefont {Retin{\`o}},\ and\ \citenamefont
		{Andr{\'e}}}]{Vaivads06}%
	\BibitemOpen
	\bibfield  {author} {\bibinfo {author} {\bibfnamefont {A.}~\bibnamefont
			{Vaivads}}, \bibinfo {author} {\bibfnamefont {A.}~\bibnamefont {Retin{\`o}}},
		\ and\ \bibinfo {author} {\bibfnamefont {M.}~\bibnamefont {Andr{\'e}}},\
	}\href {\doibase 10.1007/s11214-006-7019-3} {\bibfield  {journal} {\bibinfo
			{journal} {Space Science Reviews}\ }\textbf {\bibinfo {volume} {122}},\
		\bibinfo {pages} {19} (\bibinfo {year} {2006})}\BibitemShut {NoStop}%
	\bibitem [{\citenamefont {Phan}\ \emph {et~al.}(2006)\citenamefont {Phan},
		\citenamefont {Gosling}, \citenamefont {Davis}, \citenamefont {Skoug},
		\citenamefont {{\O}ieroset}, \citenamefont {Lin}, \citenamefont {Lepping},
		\citenamefont {McComas}, \citenamefont {Smith}, \citenamefont {Reme},\ and\
		\citenamefont {Balogh}}]{Phan06}%
	\BibitemOpen
	\bibfield  {author} {\bibinfo {author} {\bibfnamefont {T.~D.}\ \bibnamefont
			{Phan}}, \bibinfo {author} {\bibfnamefont {J.~T.}\ \bibnamefont {Gosling}},
		\bibinfo {author} {\bibfnamefont {M.~S.}\ \bibnamefont {Davis}}, \bibinfo
		{author} {\bibfnamefont {R.~M.}\ \bibnamefont {Skoug}}, \bibinfo {author}
		{\bibfnamefont {M.}~\bibnamefont {{\O}ieroset}}, \bibinfo {author}
		{\bibfnamefont {R.~P.}\ \bibnamefont {Lin}}, \bibinfo {author} {\bibfnamefont
			{R.~P.}\ \bibnamefont {Lepping}}, \bibinfo {author} {\bibfnamefont {D.~J.}\
			\bibnamefont {McComas}}, \bibinfo {author} {\bibfnamefont {C.~W.}\
			\bibnamefont {Smith}}, \bibinfo {author} {\bibfnamefont {H.}~\bibnamefont
			{Reme}}, \ and\ \bibinfo {author} {\bibfnamefont {A.}~\bibnamefont
			{Balogh}},\ }\href {http://dx.doi.org/10.1038/nature04393} {\bibfield
		{journal} {\bibinfo  {journal} {Nature}\ }\textbf {\bibinfo {volume} {439}},\
		\bibinfo {pages} {175} (\bibinfo {year} {2006})}\BibitemShut {NoStop}%
	\bibitem [{\citenamefont {Retin{\`o}}\ \emph {et~al.}(2007)\citenamefont
		{Retin{\`o}}, \citenamefont {Sundkvist}, \citenamefont {Vaivads},
		\citenamefont {Mozer}, \citenamefont {Andr{\'e}},\ and\ \citenamefont
		{Owen}}]{Retino07}%
	\BibitemOpen
	\bibfield  {author} {\bibinfo {author} {\bibfnamefont {A.}~\bibnamefont
			{Retin{\`o}}}, \bibinfo {author} {\bibfnamefont {D.}~\bibnamefont
			{Sundkvist}}, \bibinfo {author} {\bibfnamefont {A.}~\bibnamefont {Vaivads}},
		\bibinfo {author} {\bibfnamefont {F.}~\bibnamefont {Mozer}}, \bibinfo
		{author} {\bibfnamefont {M.}~\bibnamefont {Andr{\'e}}}, \ and\ \bibinfo
		{author} {\bibfnamefont {C.~J.}\ \bibnamefont {Owen}},\ }\href
	{http://dx.doi.org/10.1038/nphys574} {\bibfield  {journal} {\bibinfo
			{journal} {Nature Physics}\ }\textbf {\bibinfo {volume} {3}},\ \bibinfo
		{pages} {235} (\bibinfo {year} {2007})}\BibitemShut {NoStop}%
	\bibitem [{\citenamefont {Phan}\ \emph {et~al.}(2018)\citenamefont {Phan},
		\citenamefont {Eastwood}, \citenamefont {Shay}, \citenamefont {Drake},
		\citenamefont {Sonnerup}, \citenamefont {Fujimoto}, \citenamefont {Cassak},
		\citenamefont {{\O}ieroset}, \citenamefont {Burch}, \citenamefont {Torbert},
		\citenamefont {Rager}, \citenamefont {Dorelli}, \citenamefont {Gershman},
		\citenamefont {Pollock}, \citenamefont {Pyakurel}, \citenamefont {Haggerty},
		\citenamefont {Khotyaintsev}, \citenamefont {Lavraud}, \citenamefont {Saito},
		\citenamefont {Oka}, \citenamefont {Ergun}, \citenamefont {Retin\`o},
		\citenamefont {Le~Contel}, \citenamefont {Argall}, \citenamefont {Giles},
		\citenamefont {Moore}, \citenamefont {Wilder}, \citenamefont {Strangeway},
		\citenamefont {Russell}, \citenamefont {Lindqvist},\ and\ \citenamefont
		{Magnes}}]{Phan18}%
	\BibitemOpen
	\bibfield  {author} {\bibinfo {author} {\bibfnamefont {T.~D.}\ \bibnamefont
			{Phan}}, \bibinfo {author} {\bibfnamefont {J.~P.}\ \bibnamefont {Eastwood}},
		\bibinfo {author} {\bibfnamefont {M.~A.}\ \bibnamefont {Shay}}, \bibinfo
		{author} {\bibfnamefont {J.~F.}\ \bibnamefont {Drake}}, \bibinfo {author}
		{\bibfnamefont {B.~U.~{\"O}.}\ \bibnamefont {Sonnerup}}, \bibinfo {author}
		{\bibfnamefont {M.}~\bibnamefont {Fujimoto}}, \bibinfo {author}
		{\bibfnamefont {P.~A.}\ \bibnamefont {Cassak}}, \bibinfo {author}
		{\bibfnamefont {M.}~\bibnamefont {{\O}ieroset}}, \bibinfo {author}
		{\bibfnamefont {J.~L.}\ \bibnamefont {Burch}}, \bibinfo {author}
		{\bibfnamefont {R.~B.}\ \bibnamefont {Torbert}}, \bibinfo {author}
		{\bibfnamefont {A.~C.}\ \bibnamefont {Rager}}, \bibinfo {author}
		{\bibfnamefont {J.~C.}\ \bibnamefont {Dorelli}}, \bibinfo {author}
		{\bibfnamefont {D.~J.}\ \bibnamefont {Gershman}}, \bibinfo {author}
		{\bibfnamefont {C.}~\bibnamefont {Pollock}}, \bibinfo {author} {\bibfnamefont
			{P.~S.}\ \bibnamefont {Pyakurel}}, \bibinfo {author} {\bibfnamefont {C.~C.}\
			\bibnamefont {Haggerty}}, \bibinfo {author} {\bibfnamefont {Y.}~\bibnamefont
			{Khotyaintsev}}, \bibinfo {author} {\bibfnamefont {B.}~\bibnamefont
			{Lavraud}}, \bibinfo {author} {\bibfnamefont {Y.}~\bibnamefont {Saito}},
		\bibinfo {author} {\bibfnamefont {M.}~\bibnamefont {Oka}}, \bibinfo {author}
		{\bibfnamefont {R.~E.}\ \bibnamefont {Ergun}}, \bibinfo {author}
		{\bibfnamefont {A.}~\bibnamefont {Retin\`o}}, \bibinfo {author}
		{\bibfnamefont {O.}~\bibnamefont {Le~Contel}}, \bibinfo {author}
		{\bibfnamefont {M.~R.}\ \bibnamefont {Argall}}, \bibinfo {author}
		{\bibfnamefont {B.~L.}\ \bibnamefont {Giles}}, \bibinfo {author}
		{\bibfnamefont {T.~E.}\ \bibnamefont {Moore}}, \bibinfo {author}
		{\bibfnamefont {F.~D.}\ \bibnamefont {Wilder}}, \bibinfo {author}
		{\bibfnamefont {R.~J.}\ \bibnamefont {Strangeway}}, \bibinfo {author}
		{\bibfnamefont {C.~T.}\ \bibnamefont {Russell}}, \bibinfo {author}
		{\bibfnamefont {P.~A.}\ \bibnamefont {Lindqvist}}, \ and\ \bibinfo {author}
		{\bibfnamefont {W.}~\bibnamefont {Magnes}},\ }\href {\doibase
		10.1038/s41586-018-0091-5} {\bibfield  {journal} {\bibinfo  {journal}
			{Nature}\ }\textbf {\bibinfo {volume} {557}},\ \bibinfo {pages} {202}
		(\bibinfo {year} {2018})}\BibitemShut {NoStop}%
	\bibitem [{\citenamefont {Paschmann}\ \emph {et~al.}(1979)\citenamefont
		{Paschmann}, \citenamefont {Sonnerup}, \citenamefont {Papamastorakis},
		\citenamefont {Sckopke}, \citenamefont {Haerendel}, \citenamefont {Bame},
		\citenamefont {Asbridge}, \citenamefont {Gosling}, \citenamefont {Russell},\
		and\ \citenamefont {Elphic}}]{Paschmann79}%
	\BibitemOpen
	\bibfield  {author} {\bibinfo {author} {\bibfnamefont {G.}~\bibnamefont
			{Paschmann}}, \bibinfo {author} {\bibfnamefont {B.~U.~{\"O}.}\ \bibnamefont
			{Sonnerup}}, \bibinfo {author} {\bibfnamefont {I.}~\bibnamefont
			{Papamastorakis}}, \bibinfo {author} {\bibfnamefont {N.}~\bibnamefont
			{Sckopke}}, \bibinfo {author} {\bibfnamefont {G.}~\bibnamefont {Haerendel}},
		\bibinfo {author} {\bibfnamefont {S.~J.}\ \bibnamefont {Bame}}, \bibinfo
		{author} {\bibfnamefont {J.~R.}\ \bibnamefont {Asbridge}}, \bibinfo {author}
		{\bibfnamefont {J.~T.}\ \bibnamefont {Gosling}}, \bibinfo {author}
		{\bibfnamefont {C.~T.}\ \bibnamefont {Russell}}, \ and\ \bibinfo {author}
		{\bibfnamefont {R.~C.}\ \bibnamefont {Elphic}},\ }\href
	{http://dx.doi.org/10.1038/282243a0} {\bibfield  {journal} {\bibinfo
			{journal} {Nature}\ }\textbf {\bibinfo {volume} {282}},\ \bibinfo {pages}
		{243} (\bibinfo {year} {1979})}\BibitemShut {NoStop}%
	\bibitem [{\citenamefont {Vaivads}\ \emph {et~al.}(2004)\citenamefont
		{Vaivads}, \citenamefont {Khotyaintsev}, \citenamefont {Andr\'e},
		\citenamefont {Retin\`o}, \citenamefont {Buchert}, \citenamefont {Rogers},
		\citenamefont {D\'ecr\'eau}, \citenamefont {Paschmann},\ and\ \citenamefont
		{Phan}}]{Vaivads04}%
	\BibitemOpen
	\bibfield  {author} {\bibinfo {author} {\bibfnamefont {A.}~\bibnamefont
			{Vaivads}}, \bibinfo {author} {\bibfnamefont {Y.}~\bibnamefont
			{Khotyaintsev}}, \bibinfo {author} {\bibfnamefont {M.}~\bibnamefont
			{Andr\'e}}, \bibinfo {author} {\bibfnamefont {A.}~\bibnamefont {Retin\`o}},
		\bibinfo {author} {\bibfnamefont {S.~C.}\ \bibnamefont {Buchert}}, \bibinfo
		{author} {\bibfnamefont {B.~N.}\ \bibnamefont {Rogers}}, \bibinfo {author}
		{\bibfnamefont {P.}~\bibnamefont {D\'ecr\'eau}}, \bibinfo {author}
		{\bibfnamefont {G.}~\bibnamefont {Paschmann}}, \ and\ \bibinfo {author}
		{\bibfnamefont {T.~D.}\ \bibnamefont {Phan}},\ }\href {\doibase
		10.1103/PhysRevLett.93.105001} {\bibfield  {journal} {\bibinfo  {journal}
			{Phys. Rev. Lett.}\ }\textbf {\bibinfo {volume} {93}},\ \bibinfo {pages}
		{105001} (\bibinfo {year} {2004})}\BibitemShut {NoStop}%
	\bibitem [{\citenamefont {Burch}\ \emph
		{et~al.}(2016{\natexlab{a}})\citenamefont {Burch}, \citenamefont {Torbert},
		\citenamefont {Phan}, \citenamefont {Chen}, \citenamefont {Moore},
		\citenamefont {Ergun}, \citenamefont {Eastwood}, \citenamefont {Gershman},
		\citenamefont {Cassak}, \citenamefont {Argall}, \citenamefont {Wang},
		\citenamefont {Hesse}, \citenamefont {Pollock}, \citenamefont {Giles},
		\citenamefont {Nakamura}, \citenamefont {Mauk}, \citenamefont {Fuselier},
		\citenamefont {Russell}, \citenamefont {Strangeway}, \citenamefont {Drake},
		\citenamefont {Shay}, \citenamefont {Khotyaintsev}, \citenamefont
		{Lindqvist}, \citenamefont {Marklund}, \citenamefont {Wilder}, \citenamefont
		{Young}, \citenamefont {Torkar}, \citenamefont {Goldstein}, \citenamefont
		{Dorelli}, \citenamefont {Avanov}, \citenamefont {Oka}, \citenamefont
		{Baker}, \citenamefont {Jaynes}, \citenamefont {Goodrich}, \citenamefont
		{Cohen}, \citenamefont {Turner}, \citenamefont {Fennell}, \citenamefont
		{Blake}, \citenamefont {Clemmons}, \citenamefont {Goldman}, \citenamefont
		{Newman}, \citenamefont {Petrinec}, \citenamefont {Trattner}, \citenamefont
		{Lavraud}, \citenamefont {Reiff}, \citenamefont {Baumjohann}, \citenamefont
		{Magnes}, \citenamefont {Steller}, \citenamefont {Lewis}, \citenamefont
		{Saito}, \citenamefont {Coffey},\ and\ \citenamefont {Chandler}}]{Burch16}%
	\BibitemOpen
	\bibfield  {author} {\bibinfo {author} {\bibfnamefont {J.~L.}\ \bibnamefont
			{Burch}}, \bibinfo {author} {\bibfnamefont {R.~B.}\ \bibnamefont {Torbert}},
		\bibinfo {author} {\bibfnamefont {T.~D.}\ \bibnamefont {Phan}}, \bibinfo
		{author} {\bibfnamefont {L.-J.}\ \bibnamefont {Chen}}, \bibinfo {author}
		{\bibfnamefont {T.~E.}\ \bibnamefont {Moore}}, \bibinfo {author}
		{\bibfnamefont {R.~E.}\ \bibnamefont {Ergun}}, \bibinfo {author}
		{\bibfnamefont {J.~P.}\ \bibnamefont {Eastwood}}, \bibinfo {author}
		{\bibfnamefont {D.~J.}\ \bibnamefont {Gershman}}, \bibinfo {author}
		{\bibfnamefont {P.~A.}\ \bibnamefont {Cassak}}, \bibinfo {author}
		{\bibfnamefont {M.~R.}\ \bibnamefont {Argall}}, \bibinfo {author}
		{\bibfnamefont {S.}~\bibnamefont {Wang}}, \bibinfo {author} {\bibfnamefont
			{M.}~\bibnamefont {Hesse}}, \bibinfo {author} {\bibfnamefont {C.~J.}\
			\bibnamefont {Pollock}}, \bibinfo {author} {\bibfnamefont {B.~L.}\
			\bibnamefont {Giles}}, \bibinfo {author} {\bibfnamefont {R.}~\bibnamefont
			{Nakamura}}, \bibinfo {author} {\bibfnamefont {B.~H.}\ \bibnamefont {Mauk}},
		\bibinfo {author} {\bibfnamefont {S.~A.}\ \bibnamefont {Fuselier}}, \bibinfo
		{author} {\bibfnamefont {C.~T.}\ \bibnamefont {Russell}}, \bibinfo {author}
		{\bibfnamefont {R.~J.}\ \bibnamefont {Strangeway}}, \bibinfo {author}
		{\bibfnamefont {J.~F.}\ \bibnamefont {Drake}}, \bibinfo {author}
		{\bibfnamefont {M.~A.}\ \bibnamefont {Shay}}, \bibinfo {author}
		{\bibfnamefont {Y.~V.}\ \bibnamefont {Khotyaintsev}}, \bibinfo {author}
		{\bibfnamefont {P.-A.}\ \bibnamefont {Lindqvist}}, \bibinfo {author}
		{\bibfnamefont {G.}~\bibnamefont {Marklund}}, \bibinfo {author}
		{\bibfnamefont {F.~D.}\ \bibnamefont {Wilder}}, \bibinfo {author}
		{\bibfnamefont {D.~T.}\ \bibnamefont {Young}}, \bibinfo {author}
		{\bibfnamefont {K.}~\bibnamefont {Torkar}}, \bibinfo {author} {\bibfnamefont
			{J.}~\bibnamefont {Goldstein}}, \bibinfo {author} {\bibfnamefont {J.~C.}\
			\bibnamefont {Dorelli}}, \bibinfo {author} {\bibfnamefont {L.~A.}\
			\bibnamefont {Avanov}}, \bibinfo {author} {\bibfnamefont {M.}~\bibnamefont
			{Oka}}, \bibinfo {author} {\bibfnamefont {D.~N.}\ \bibnamefont {Baker}},
		\bibinfo {author} {\bibfnamefont {A.~N.}\ \bibnamefont {Jaynes}}, \bibinfo
		{author} {\bibfnamefont {K.~A.}\ \bibnamefont {Goodrich}}, \bibinfo {author}
		{\bibfnamefont {I.~J.}\ \bibnamefont {Cohen}}, \bibinfo {author}
		{\bibfnamefont {D.~L.}\ \bibnamefont {Turner}}, \bibinfo {author}
		{\bibfnamefont {J.~F.}\ \bibnamefont {Fennell}}, \bibinfo {author}
		{\bibfnamefont {J.~B.}\ \bibnamefont {Blake}}, \bibinfo {author}
		{\bibfnamefont {J.}~\bibnamefont {Clemmons}}, \bibinfo {author}
		{\bibfnamefont {M.}~\bibnamefont {Goldman}}, \bibinfo {author} {\bibfnamefont
			{D.}~\bibnamefont {Newman}}, \bibinfo {author} {\bibfnamefont {S.~M.}\
			\bibnamefont {Petrinec}}, \bibinfo {author} {\bibfnamefont {K.~J.}\
			\bibnamefont {Trattner}}, \bibinfo {author} {\bibfnamefont {B.}~\bibnamefont
			{Lavraud}}, \bibinfo {author} {\bibfnamefont {P.~H.}\ \bibnamefont {Reiff}},
		\bibinfo {author} {\bibfnamefont {W.}~\bibnamefont {Baumjohann}}, \bibinfo
		{author} {\bibfnamefont {W.}~\bibnamefont {Magnes}}, \bibinfo {author}
		{\bibfnamefont {M.}~\bibnamefont {Steller}}, \bibinfo {author} {\bibfnamefont
			{W.}~\bibnamefont {Lewis}}, \bibinfo {author} {\bibfnamefont
			{Y.}~\bibnamefont {Saito}}, \bibinfo {author} {\bibfnamefont
			{V.}~\bibnamefont {Coffey}}, \ and\ \bibinfo {author} {\bibfnamefont
			{M.}~\bibnamefont {Chandler}},\ }\href {\doibase 10.1126/science.aaf2939}
	{\bibfield  {journal} {\bibinfo  {journal} {Science}\ }\textbf {\bibinfo
			{volume} {352}} (\bibinfo {year} {2016}{\natexlab{a}}),\
		10.1126/science.aaf2939}\BibitemShut {NoStop}%
	\bibitem [{\citenamefont {{\O}ieroset}\ \emph {et~al.}(2001)\citenamefont
		{{\O}ieroset}, \citenamefont {Phan}, \citenamefont {Fujimoto}, \citenamefont
		{Lin},\ and\ \citenamefont {Lepping}}]{Oieroset01}%
	\BibitemOpen
	\bibfield  {author} {\bibinfo {author} {\bibfnamefont {M.}~\bibnamefont
			{{\O}ieroset}}, \bibinfo {author} {\bibfnamefont {T.~D.}\ \bibnamefont
			{Phan}}, \bibinfo {author} {\bibfnamefont {M.}~\bibnamefont {Fujimoto}},
		\bibinfo {author} {\bibfnamefont {R.~P.}\ \bibnamefont {Lin}}, \ and\
		\bibinfo {author} {\bibfnamefont {R.~P.}\ \bibnamefont {Lepping}},\ }\href
	{http://dx.doi.org/10.1038/35086520} {\bibfield  {journal} {\bibinfo
			{journal} {Nature}\ }\textbf {\bibinfo {volume} {412}},\ \bibinfo {pages}
		{414} (\bibinfo {year} {2001})}\BibitemShut {NoStop}%
	\bibitem [{\citenamefont {Pritchett}(2008)}]{Pritchett08}%
	\BibitemOpen
	\bibfield  {author} {\bibinfo {author} {\bibfnamefont {P.~L.}\ \bibnamefont
			{Pritchett}},\ }\href
	{https://agupubs.onlinelibrary.wiley.com/doi/abs/10.1029/2007JA012930}
	{\bibfield  {journal} {\bibinfo  {journal} {Journal of Geophysical Research:
				Space Physics}\ }\textbf {\bibinfo {volume} {113}} (\bibinfo {year}
		{2008})}\BibitemShut {NoStop}%
	\bibitem [{\citenamefont {Mozer}\ \emph {et~al.}(2003)\citenamefont {Mozer},
		\citenamefont {Bale}, \citenamefont {Phan},\ and\ \citenamefont
		{Osborne}}]{Mozer03}%
	\BibitemOpen
	\bibfield  {author} {\bibinfo {author} {\bibfnamefont {F.~S.}\ \bibnamefont
			{Mozer}}, \bibinfo {author} {\bibfnamefont {S.~D.}\ \bibnamefont {Bale}},
		\bibinfo {author} {\bibfnamefont {T.~D.}\ \bibnamefont {Phan}}, \ and\
		\bibinfo {author} {\bibfnamefont {J.~A.}\ \bibnamefont {Osborne}},\ }\href
	{\doibase 10.1103/PhysRevLett.91.245002} {\bibfield  {journal} {\bibinfo
			{journal} {Phys. Rev. Lett.}\ }\textbf {\bibinfo {volume} {91}},\ \bibinfo
		{pages} {245002} (\bibinfo {year} {2003})}\BibitemShut {NoStop}%
	\bibitem [{\citenamefont {Mozer}\ \emph {et~al.}(2005)\citenamefont {Mozer},
		\citenamefont {Bale}, \citenamefont {McFadden},\ and\ \citenamefont
		{Torbert}}]{Mozer05}%
	\BibitemOpen
	\bibfield  {author} {\bibinfo {author} {\bibfnamefont {F.~S.}\ \bibnamefont
			{Mozer}}, \bibinfo {author} {\bibfnamefont {S.~D.}\ \bibnamefont {Bale}},
		\bibinfo {author} {\bibfnamefont {J.~P.}\ \bibnamefont {McFadden}}, \ and\
		\bibinfo {author} {\bibfnamefont {R.~B.}\ \bibnamefont {Torbert}},\ }\href
	{https://agupubs.onlinelibrary.wiley.com/doi/abs/10.1029/2005GL024092}
	{\bibfield  {journal} {\bibinfo  {journal} {Geophysical Research Letters}\
		}\textbf {\bibinfo {volume} {32}},\ \bibinfo {pages} {24102} (\bibinfo {year}
		{2005})}\BibitemShut {NoStop}%
	\bibitem [{\citenamefont {Shay}\ \emph {et~al.}(2016)\citenamefont {Shay},
		\citenamefont {Phan}, \citenamefont {Haggerty}, \citenamefont {Fujimoto},
		\citenamefont {Drake}, \citenamefont {Malakit}, \citenamefont {Cassak},\ and\
		\citenamefont {Swisdak}}]{Shay16}%
	\BibitemOpen
	\bibfield  {author} {\bibinfo {author} {\bibfnamefont {M.~A.}\ \bibnamefont
			{Shay}}, \bibinfo {author} {\bibfnamefont {T.~D.}\ \bibnamefont {Phan}},
		\bibinfo {author} {\bibfnamefont {C.~C.}\ \bibnamefont {Haggerty}}, \bibinfo
		{author} {\bibfnamefont {M.}~\bibnamefont {Fujimoto}}, \bibinfo {author}
		{\bibfnamefont {J.~F.}\ \bibnamefont {Drake}}, \bibinfo {author}
		{\bibfnamefont {K.}~\bibnamefont {Malakit}}, \bibinfo {author} {\bibfnamefont
			{P.~A.}\ \bibnamefont {Cassak}}, \ and\ \bibinfo {author} {\bibfnamefont
			{M.}~\bibnamefont {Swisdak}},\ }\href {\doibase 10.1002/2016GL069034}
	{\bibfield  {journal} {\bibinfo  {journal} {Geophysical Research Letters}\
		}\textbf {\bibinfo {volume} {43}},\ \bibinfo {pages} {4145} (\bibinfo {year}
		{2016})}\BibitemShut {NoStop}%
		\bibitem [{\citenamefont {Q}(2016)}]{Q}%
	\BibitemOpen
	\bibfield aThe electron agyrotropy $\sqrt{Q}$ is defined as $Q = (P_{e,xy}^2 +  P_{e,xz}^2 +  P_{e,yz}^2)/(P_{e,\perp}^2 + 2P_{e,\perp}P_{e,||})$, where $P_{e,ij}$ with $i,j= x,y,z$ are the out of diagonal elements and $P_{e,||}$ and $P_{e,\perp}$ are the diagonal elements of the electron pressure tensor $\overline{P}_e$ written using a set formed by a magnetic field-aligned unit vector and two vectors orthogonal to the magnetic field. Gyrotropic tensors have $\sqrt{Q} = 0$, while maximal departures from gyrotropy imply $\sqrt{Q} = 1$ \BibitemShut
	{NoStop}%
	\bibitem [{\citenamefont {Swisdak}(2016)}]{Swisdak16}%
	\BibitemOpen
	\bibfield  {author} {\bibinfo {author} {\bibfnamefont {M.}~\bibnamefont
			{Swisdak}},\ }\href {\doibase 10.1002/2015GL066980} {\bibfield  {journal}
		{\bibinfo  {journal} {Geophysical Research Letters}\ }\textbf {\bibinfo
			{volume} {43}},\ \bibinfo {pages} {43} (\bibinfo {year} {2016})}\BibitemShut
	{NoStop}%
	\bibitem [{\citenamefont {Zenitani}\ \emph {et~al.}(2011)\citenamefont
		{Zenitani}, \citenamefont {Hesse}, \citenamefont {Klimas},\ and\
		\citenamefont {Kuznetsova}}]{Zenitani11}%
	\BibitemOpen
	\bibfield  {author} {\bibinfo {author} {\bibfnamefont {S.}~\bibnamefont
			{Zenitani}}, \bibinfo {author} {\bibfnamefont {M.}~\bibnamefont {Hesse}},
		\bibinfo {author} {\bibfnamefont {A.}~\bibnamefont {Klimas}}, \ and\ \bibinfo
		{author} {\bibfnamefont {M.}~\bibnamefont {Kuznetsova}},\ }\href {\doibase
		10.1103/PhysRevLett.106.195003} {\bibfield  {journal} {\bibinfo  {journal}
			{Phys. Rev. Lett.}\ }\textbf {\bibinfo {volume} {106}},\ \bibinfo {pages}
		{195003} (\bibinfo {year} {2011})}\BibitemShut {NoStop}%
	\bibitem [{\citenamefont {Hesse}\ \emph {et~al.}(2014)\citenamefont {Hesse},
		\citenamefont {Aunai}, \citenamefont {Sibeck},\ and\ \citenamefont
		{Birn}}]{Hesse14}%
	\BibitemOpen
	\bibfield  {author} {\bibinfo {author} {\bibfnamefont {M.}~\bibnamefont
			{Hesse}}, \bibinfo {author} {\bibfnamefont {N.}~\bibnamefont {Aunai}},
		\bibinfo {author} {\bibfnamefont {D.}~\bibnamefont {Sibeck}}, \ and\ \bibinfo
		{author} {\bibfnamefont {J.}~\bibnamefont {Birn}},\ }\href {\doibase
		10.1002/2014GL061586} {\bibfield  {journal} {\bibinfo  {journal} {Geophysical
				Research Letters}\ }\textbf {\bibinfo {volume} {41}},\ \bibinfo {pages}
		{8673} (\bibinfo {year} {2014})}\BibitemShut {NoStop}%
	\bibitem [{\citenamefont {Bessho}\ \emph {et~al.}(2016)\citenamefont {Bessho},
		\citenamefont {Chen},\ and\ \citenamefont {Hesse}}]{Bessho16}%
	\BibitemOpen
	\bibfield  {author} {\bibinfo {author} {\bibfnamefont {N.}~\bibnamefont
			{Bessho}}, \bibinfo {author} {\bibfnamefont {L.-J.}\ \bibnamefont {Chen}}, \
		and\ \bibinfo {author} {\bibfnamefont {M.}~\bibnamefont {Hesse}},\ }\href
	{\doibase 10.1002/2016GL067886} {\bibfield  {journal} {\bibinfo  {journal}
			{Geophysical Research Letters}\ }\textbf {\bibinfo {volume} {43}},\ \bibinfo
		{pages} {1828} (\bibinfo {year} {2016})}\BibitemShut {NoStop}%
	\bibitem [{\citenamefont {Lapenta}\ \emph {et~al.}(2017)\citenamefont
		{Lapenta}, \citenamefont {Berchem}, \citenamefont {Zhou}, \citenamefont
		{Walker}, \citenamefont {El-Alaoui}, \citenamefont {Goldstein}, \citenamefont
		{Paterson}, \citenamefont {Giles}, \citenamefont {Pollock}, \citenamefont
		{Russell}, \citenamefont {Strangeway}, \citenamefont {Ergun}, \citenamefont
		{Khotyaintsev}, \citenamefont {Torbert},\ and\ \citenamefont
		{Burch}}]{Lapenta17}%
	\BibitemOpen
	\bibfield  {author} {\bibinfo {author} {\bibfnamefont {G.}~\bibnamefont
			{Lapenta}}, \bibinfo {author} {\bibfnamefont {J.}~\bibnamefont {Berchem}},
		\bibinfo {author} {\bibfnamefont {M.}~\bibnamefont {Zhou}}, \bibinfo {author}
		{\bibfnamefont {R.~J.}\ \bibnamefont {Walker}}, \bibinfo {author}
		{\bibfnamefont {M.}~\bibnamefont {El-Alaoui}}, \bibinfo {author}
		{\bibfnamefont {M.~L.}\ \bibnamefont {Goldstein}}, \bibinfo {author}
		{\bibfnamefont {W.~R.}\ \bibnamefont {Paterson}}, \bibinfo {author}
		{\bibfnamefont {B.~L.}\ \bibnamefont {Giles}}, \bibinfo {author}
		{\bibfnamefont {C.~J.}\ \bibnamefont {Pollock}}, \bibinfo {author}
		{\bibfnamefont {C.~T.}\ \bibnamefont {Russell}}, \bibinfo {author}
		{\bibfnamefont {R.~J.}\ \bibnamefont {Strangeway}}, \bibinfo {author}
		{\bibfnamefont {R.~E.}\ \bibnamefont {Ergun}}, \bibinfo {author}
		{\bibfnamefont {Y.~V.}\ \bibnamefont {Khotyaintsev}}, \bibinfo {author}
		{\bibfnamefont {R.~B.}\ \bibnamefont {Torbert}}, \ and\ \bibinfo {author}
		{\bibfnamefont {J.~L.}\ \bibnamefont {Burch}},\ }\href {\doibase
		10.1002/2016JA023290} {\bibfield  {journal} {\bibinfo  {journal} {Journal of
				Geophysical Research: Space Physics}\ }\textbf {\bibinfo {volume} {122}},\
		\bibinfo {pages} {2024} (\bibinfo {year} {2017})}\BibitemShut {NoStop}%
	\bibitem [{\citenamefont {Egedal}\ \emph {et~al.}(2018)\citenamefont {Egedal},
		\citenamefont {Le}, \citenamefont {Daughton}, \citenamefont {Wetherton},
		\citenamefont {Cassak}, \citenamefont {Burch}, \citenamefont {Lavraud},
		\citenamefont {Dorelli}, \citenamefont {Gershman},\ and\ \citenamefont
		{Avanov}}]{Egedal18}%
	\BibitemOpen
	\bibfield  {author} {\bibinfo {author} {\bibfnamefont {J.}~\bibnamefont
			{Egedal}}, \bibinfo {author} {\bibfnamefont {A.}~\bibnamefont {Le}}, \bibinfo
		{author} {\bibfnamefont {W.}~\bibnamefont {Daughton}}, \bibinfo {author}
		{\bibfnamefont {B.}~\bibnamefont {Wetherton}}, \bibinfo {author}
		{\bibfnamefont {P.~A.}\ \bibnamefont {Cassak}}, \bibinfo {author}
		{\bibfnamefont {J.~L.}\ \bibnamefont {Burch}}, \bibinfo {author}
		{\bibfnamefont {B.}~\bibnamefont {Lavraud}}, \bibinfo {author} {\bibfnamefont
			{J.}~\bibnamefont {Dorelli}}, \bibinfo {author} {\bibfnamefont {D.~J.}\
			\bibnamefont {Gershman}}, \ and\ \bibinfo {author} {\bibfnamefont {L.~A.}\
			\bibnamefont {Avanov}},\ }\href {\doibase 10.1103/PhysRevLett.120.055101}
	{\bibfield  {journal} {\bibinfo  {journal} {Phys. Rev. Lett.}\ }\textbf
		{\bibinfo {volume} {120}},\ \bibinfo {pages} {055101} (\bibinfo {year}
		{2018})}\BibitemShut {NoStop}%
	\bibitem [{\citenamefont {Burch}\ \emph
		{et~al.}(2016{\natexlab{b}})\citenamefont {Burch}, \citenamefont {Moore},
		\citenamefont {Torbert},\ and\ \citenamefont {Giles}}]{Burch16a}%
	\BibitemOpen
	\bibfield  {author} {\bibinfo {author} {\bibfnamefont {J.~L.}\ \bibnamefont
			{Burch}}, \bibinfo {author} {\bibfnamefont {T.~E.}\ \bibnamefont {Moore}},
		\bibinfo {author} {\bibfnamefont {R.~B.}\ \bibnamefont {Torbert}}, \ and\
		\bibinfo {author} {\bibfnamefont {B.~L.}\ \bibnamefont {Giles}},\ }\href
	{\doibase 10.1007/s11214-015-0164-9} {\bibfield  {journal} {\bibinfo
			{journal} {Space Science Reviews}\ }\textbf {\bibinfo {volume} {199}},\
		\bibinfo {pages} {5} (\bibinfo {year} {2016}{\natexlab{b}})}\BibitemShut
	{NoStop}%
	\bibitem [{\citenamefont {Webster}\ \emph {et~al.}(2018)\citenamefont
		{Webster}, \citenamefont {Burch}, \citenamefont {Reiff}, \citenamefont
		{Daou}, \citenamefont {Genestreti}, \citenamefont {Graham}, \citenamefont
		{Torbert}, \citenamefont {Ergun}, \citenamefont {Sazykin}, \citenamefont
		{Marshall}, \citenamefont {Allen}, \citenamefont {Chen}, \citenamefont
		{Wang}, \citenamefont {Phan}, \citenamefont {Giles}, \citenamefont {Moore},
		\citenamefont {Fuselier}, \citenamefont {Cozzani}, \citenamefont {Russell},
		\citenamefont {Eriksson}, \citenamefont {Rager}, \citenamefont {Broll},
		\citenamefont {Goodrich},\ and\ \citenamefont {Wilder}}]{Webster18}%
	\BibitemOpen
	\bibfield  {author} {\bibinfo {author} {\bibfnamefont {J.~M.}\ \bibnamefont
			{Webster}}, \bibinfo {author} {\bibfnamefont {J.~L.}\ \bibnamefont {Burch}},
		\bibinfo {author} {\bibfnamefont {P.~H.}\ \bibnamefont {Reiff}}, \bibinfo
		{author} {\bibfnamefont {A.~G.}\ \bibnamefont {Daou}}, \bibinfo {author}
		{\bibfnamefont {K.~J.}\ \bibnamefont {Genestreti}}, \bibinfo {author}
		{\bibfnamefont {D.~B.}\ \bibnamefont {Graham}}, \bibinfo {author}
		{\bibfnamefont {R.~B.}\ \bibnamefont {Torbert}}, \bibinfo {author}
		{\bibfnamefont {R.~E.}\ \bibnamefont {Ergun}}, \bibinfo {author}
		{\bibfnamefont {S.~Y.}\ \bibnamefont {Sazykin}}, \bibinfo {author}
		{\bibfnamefont {A.}~\bibnamefont {Marshall}}, \bibinfo {author}
		{\bibfnamefont {R.~C.}\ \bibnamefont {Allen}}, \bibinfo {author}
		{\bibfnamefont {L.-J.}\ \bibnamefont {Chen}}, \bibinfo {author}
		{\bibfnamefont {S.}~\bibnamefont {Wang}}, \bibinfo {author} {\bibfnamefont
			{T.~D.}\ \bibnamefont {Phan}}, \bibinfo {author} {\bibfnamefont {B.~L.}\
			\bibnamefont {Giles}}, \bibinfo {author} {\bibfnamefont {T.~E.}\ \bibnamefont
			{Moore}}, \bibinfo {author} {\bibfnamefont {S.~A.}\ \bibnamefont {Fuselier}},
		\bibinfo {author} {\bibfnamefont {G.}~\bibnamefont {Cozzani}}, \bibinfo
		{author} {\bibfnamefont {C.~T.}\ \bibnamefont {Russell}}, \bibinfo {author}
		{\bibfnamefont {S.}~\bibnamefont {Eriksson}}, \bibinfo {author}
		{\bibfnamefont {A.~C.}\ \bibnamefont {Rager}}, \bibinfo {author}
		{\bibfnamefont {J.~M.}\ \bibnamefont {Broll}}, \bibinfo {author}
		{\bibfnamefont {K.}~\bibnamefont {Goodrich}}, \ and\ \bibinfo {author}
		{\bibfnamefont {F.}~\bibnamefont {Wilder}},\ }\href
	{https://agupubs.onlinelibrary.wiley.com/doi/abs/10.1029/2018JA025245}
	{\bibfield  {journal} {\bibinfo  {journal} {Journal of Geophysical Research:
				Space Physics}\ }\textbf {\bibinfo {volume} {123}},\ \bibinfo {pages} {4858}
		(\bibinfo {year} {2018})}\BibitemShut {NoStop}%
	\bibitem [{\citenamefont {Norgren}\ \emph {et~al.}(2016)\citenamefont
		{Norgren}, \citenamefont {Graham}, \citenamefont {Khotyaintsev},
		\citenamefont {Andr\'e}, \citenamefont {Vaivads}, \citenamefont {Chen},
		\citenamefont {Lindqvist}, \citenamefont {Marklund}, \citenamefont {Ergun},
		\citenamefont {Magnes}, \citenamefont {Strangeway}, \citenamefont {Russell},
		\citenamefont {Torbert}, \citenamefont {Paterson}, \citenamefont {Gershman},
		\citenamefont {Dorelli}, \citenamefont {Avanov}, \citenamefont {Lavraud},
		\citenamefont {Saito}, \citenamefont {Giles}, \citenamefont {Pollock},\ and\
		\citenamefont {Burch}}]{Norgren16}%
	\BibitemOpen
	\bibfield  {author} {\bibinfo {author} {\bibfnamefont {C.}~\bibnamefont
			{Norgren}}, \bibinfo {author} {\bibfnamefont {D.~B.}\ \bibnamefont {Graham}},
		\bibinfo {author} {\bibfnamefont {Y.~V.}\ \bibnamefont {Khotyaintsev}},
		\bibinfo {author} {\bibfnamefont {M.}~\bibnamefont {Andr\'e}}, \bibinfo
		{author} {\bibfnamefont {A.}~\bibnamefont {Vaivads}}, \bibinfo {author}
		{\bibfnamefont {L.-J.}\ \bibnamefont {Chen}}, \bibinfo {author}
		{\bibfnamefont {P.-A.}\ \bibnamefont {Lindqvist}}, \bibinfo {author}
		{\bibfnamefont {G.~T.}\ \bibnamefont {Marklund}}, \bibinfo {author}
		{\bibfnamefont {R.~E.}\ \bibnamefont {Ergun}}, \bibinfo {author}
		{\bibfnamefont {W.}~\bibnamefont {Magnes}}, \bibinfo {author} {\bibfnamefont
			{R.~J.}\ \bibnamefont {Strangeway}}, \bibinfo {author} {\bibfnamefont
			{C.~T.}\ \bibnamefont {Russell}}, \bibinfo {author} {\bibfnamefont {R.~B.}\
			\bibnamefont {Torbert}}, \bibinfo {author} {\bibfnamefont {W.~R.}\
			\bibnamefont {Paterson}}, \bibinfo {author} {\bibfnamefont {D.~J.}\
			\bibnamefont {Gershman}}, \bibinfo {author} {\bibfnamefont {J.~C.}\
			\bibnamefont {Dorelli}}, \bibinfo {author} {\bibfnamefont {L.~A.}\
			\bibnamefont {Avanov}}, \bibinfo {author} {\bibfnamefont {B.}~\bibnamefont
			{Lavraud}}, \bibinfo {author} {\bibfnamefont {Y.}~\bibnamefont {Saito}},
		\bibinfo {author} {\bibfnamefont {B.~L.}\ \bibnamefont {Giles}}, \bibinfo
		{author} {\bibfnamefont {C.~J.}\ \bibnamefont {Pollock}}, \ and\ \bibinfo
		{author} {\bibfnamefont {J.~L.}\ \bibnamefont {Burch}},\ }\href {\doibase
		10.1002/2016GL069205} {\bibfield  {journal} {\bibinfo  {journal} {Geophysical
				Research Letters}\ }\textbf {\bibinfo {volume} {43}},\ \bibinfo {pages}
		{6724} (\bibinfo {year} {2016})}\BibitemShut {NoStop}%
	\bibitem [{\citenamefont {Chen}\ \emph {et~al.}(2017)\citenamefont {Chen},
		\citenamefont {Hesse}, \citenamefont {Wang}, \citenamefont {Gershman},
		\citenamefont {Ergun}, \citenamefont {Burch}, \citenamefont {Bessho},
		\citenamefont {Torbert}, \citenamefont {Giles}, \citenamefont {Webster},
		\citenamefont {Pollock}, \citenamefont {Dorelli}, \citenamefont {Moore},
		\citenamefont {Paterson}, \citenamefont {Lavraud}, \citenamefont
		{Strangeway}, \citenamefont {Russell}, \citenamefont {Khotyaintsev},
		\citenamefont {Lindqvist},\ and\ \citenamefont {Avanov}}]{Chen17}%
	\BibitemOpen
	\bibfield  {author} {\bibinfo {author} {\bibfnamefont {L.-J.}\ \bibnamefont
			{Chen}}, \bibinfo {author} {\bibfnamefont {M.}~\bibnamefont {Hesse}},
		\bibinfo {author} {\bibfnamefont {S.}~\bibnamefont {Wang}}, \bibinfo {author}
		{\bibfnamefont {D.}~\bibnamefont {Gershman}}, \bibinfo {author}
		{\bibfnamefont {R.~E.}\ \bibnamefont {Ergun}}, \bibinfo {author}
		{\bibfnamefont {J.}~\bibnamefont {Burch}}, \bibinfo {author} {\bibfnamefont
			{N.}~\bibnamefont {Bessho}}, \bibinfo {author} {\bibfnamefont {R.~B.}\
			\bibnamefont {Torbert}}, \bibinfo {author} {\bibfnamefont {B.}~\bibnamefont
			{Giles}}, \bibinfo {author} {\bibfnamefont {J.}~\bibnamefont {Webster}},
		\bibinfo {author} {\bibfnamefont {C.}~\bibnamefont {Pollock}}, \bibinfo
		{author} {\bibfnamefont {J.}~\bibnamefont {Dorelli}}, \bibinfo {author}
		{\bibfnamefont {T.}~\bibnamefont {Moore}}, \bibinfo {author} {\bibfnamefont
			{W.}~\bibnamefont {Paterson}}, \bibinfo {author} {\bibfnamefont
			{B.}~\bibnamefont {Lavraud}}, \bibinfo {author} {\bibfnamefont
			{R.}~\bibnamefont {Strangeway}}, \bibinfo {author} {\bibfnamefont
			{C.}~\bibnamefont {Russell}}, \bibinfo {author} {\bibfnamefont
			{Y.}~\bibnamefont {Khotyaintsev}}, \bibinfo {author} {\bibfnamefont {P.-A.}\
			\bibnamefont {Lindqvist}}, \ and\ \bibinfo {author} {\bibfnamefont
			{L.}~\bibnamefont {Avanov}},\ }\href {\doibase 10.1002/2017JA024004}
	{\bibfield  {journal} {\bibinfo  {journal} {Journal of Geophysical Research:
				Space Physics}\ }\textbf {\bibinfo {volume} {122}},\ \bibinfo {pages} {5235}
		(\bibinfo {year} {2017})}\BibitemShut {NoStop}%
	\bibitem [{\citenamefont {Matthaeus}(1982)}]{Matthaeus82}%
	\BibitemOpen
	\bibfield  {author} {\bibinfo {author} {\bibfnamefont {W.~H.}\ \bibnamefont
			{Matthaeus}},\ }\href {\doibase 10.1029/GL009i006p00660} {\bibfield
		{journal} {\bibinfo  {journal} {Geophysical Research Letters}\ }\textbf
		{\bibinfo {volume} {9}},\ \bibinfo {pages} {660} (\bibinfo {year}
		{1982})}\BibitemShut {NoStop}%
	\bibitem [{\citenamefont {Che}\ \emph {et~al.}(2011)\citenamefont {Che},
		\citenamefont {Drake},\ and\ \citenamefont {Swisdak}}]{Che11}%
	\BibitemOpen
	\bibfield  {author} {\bibinfo {author} {\bibfnamefont {H.}~\bibnamefont
			{Che}}, \bibinfo {author} {\bibfnamefont {J.~F.}\ \bibnamefont {Drake}}, \
		and\ \bibinfo {author} {\bibfnamefont {M.}~\bibnamefont {Swisdak}},\ }\href
	{\doibase 10.1038/nature10091} {\bibfield  {journal} {\bibinfo  {journal}
			{Nature}\ }\textbf {\bibinfo {volume} {474}},\ \bibinfo {pages} {184}
		(\bibinfo {year} {2011})}\BibitemShut {NoStop}%
	\bibitem [{\citenamefont {Daughton}\ \emph {et~al.}(2011)\citenamefont
		{Daughton}, \citenamefont {Roytershteyn}, \citenamefont {Karimabadi},
		\citenamefont {Yin}, \citenamefont {Albright}, \citenamefont {Bergen},\ and\
		\citenamefont {Bowers}}]{Daughton11}%
	\BibitemOpen
	\bibfield  {author} {\bibinfo {author} {\bibfnamefont {W.}~\bibnamefont
			{Daughton}}, \bibinfo {author} {\bibfnamefont {V.}~\bibnamefont
			{Roytershteyn}}, \bibinfo {author} {\bibfnamefont {H.}~\bibnamefont
			{Karimabadi}}, \bibinfo {author} {\bibfnamefont {L.}~\bibnamefont {Yin}},
		\bibinfo {author} {\bibfnamefont {B.~J.}\ \bibnamefont {Albright}}, \bibinfo
		{author} {\bibfnamefont {B.}~\bibnamefont {Bergen}}, \ and\ \bibinfo {author}
		{\bibfnamefont {K.~J.}\ \bibnamefont {Bowers}},\ }\href {\doibase
		10.1038/nphys1965} {\bibfield  {journal} {\bibinfo  {journal} {Nat Phys}\
		}\textbf {\bibinfo {volume} {7}},\ \bibinfo {pages} {539} (\bibinfo {year}
		{2011})}\BibitemShut {NoStop}%
	\bibitem [{\citenamefont {Price}\ \emph {et~al.}(2016)\citenamefont {Price},
		\citenamefont {Swisdak}, \citenamefont {Drake}, \citenamefont {Cassak},
		\citenamefont {Dahlin},\ and\ \citenamefont {Ergun}}]{Price16}%
	\BibitemOpen
	\bibfield  {author} {\bibinfo {author} {\bibfnamefont {L.}~\bibnamefont
			{Price}}, \bibinfo {author} {\bibfnamefont {M.}~\bibnamefont {Swisdak}},
		\bibinfo {author} {\bibfnamefont {J.~F.}\ \bibnamefont {Drake}}, \bibinfo
		{author} {\bibfnamefont {P.~A.}\ \bibnamefont {Cassak}}, \bibinfo {author}
		{\bibfnamefont {J.~T.}\ \bibnamefont {Dahlin}}, \ and\ \bibinfo {author}
		{\bibfnamefont {R.~E.}\ \bibnamefont {Ergun}},\ }\href {\doibase
		10.1002/2016GL069578} {\bibfield  {journal} {\bibinfo  {journal} {Geophysical
				Research Letters}\ }\textbf {\bibinfo {volume} {43}},\ \bibinfo {pages}
		{6020} (\bibinfo {year} {2016})}\BibitemShut {NoStop}%
	\bibitem [{\citenamefont {Price}\ \emph {et~al.}(2017)\citenamefont {Price},
		\citenamefont {Swisdak}, \citenamefont {Drake}, \citenamefont {Burch},
		\citenamefont {Cassak},\ and\ \citenamefont {Ergun}}]{Price17}%
	\BibitemOpen
	\bibfield  {author} {\bibinfo {author} {\bibfnamefont {L.}~\bibnamefont
			{Price}}, \bibinfo {author} {\bibfnamefont {M.}~\bibnamefont {Swisdak}},
		\bibinfo {author} {\bibfnamefont {J.~F.}\ \bibnamefont {Drake}}, \bibinfo
		{author} {\bibfnamefont {J.~L.}\ \bibnamefont {Burch}}, \bibinfo {author}
		{\bibfnamefont {P.~A.}\ \bibnamefont {Cassak}}, \ and\ \bibinfo {author}
		{\bibfnamefont {R.~E.}\ \bibnamefont {Ergun}},\ }\href
	{https://agupubs.onlinelibrary.wiley.com/doi/abs/10.1002/2017JA024227}
	{\bibfield  {journal} {\bibinfo  {journal} {Journal of Geophysical Research:
				Space Physics}\ }\textbf {\bibinfo {volume} {122}},\ \bibinfo {pages}
		{11,086} (\bibinfo {year} {2017})}\BibitemShut {NoStop}%
	\bibitem [{\citenamefont {{Jara-Almonte}}\ \emph {et~al.}(2014)\citenamefont
		{{Jara-Almonte}}, \citenamefont {{Daughton}},\ and\ \citenamefont
		{{Ji}}}]{JaraAlmonte14}%
	\BibitemOpen
	\bibfield  {author} {\bibinfo {author} {\bibfnamefont {J.}~\bibnamefont
			{{Jara-Almonte}}}, \bibinfo {author} {\bibfnamefont {W.}~\bibnamefont
			{{Daughton}}}, \ and\ \bibinfo {author} {\bibfnamefont {H.}~\bibnamefont
			{{Ji}}},\ }\href {\doibase 10.1063/1.4867868} {\bibfield  {journal} {\bibinfo
			{journal} {Physics of Plasmas}\ }\textbf {\bibinfo {volume} {21}},\ \bibinfo
		{eid} {032114} (\bibinfo {year} {2014})}\BibitemShut {NoStop}%
	\bibitem [{\citenamefont {Swisdak}\ \emph {et~al.}(2018)\citenamefont
		{Swisdak}, \citenamefont {Drake}, \citenamefont {Price}, \citenamefont
		{Burch}, \citenamefont {Cassak},\ and\ \citenamefont {Phan}}]{Swisdak18}%
	\BibitemOpen
	\bibfield  {author} {\bibinfo {author} {\bibfnamefont {M.}~\bibnamefont
			{Swisdak}}, \bibinfo {author} {\bibfnamefont {J.~F.}\ \bibnamefont {Drake}},
		\bibinfo {author} {\bibfnamefont {L.}~\bibnamefont {Price}}, \bibinfo
		{author} {\bibfnamefont {J.~L.}\ \bibnamefont {Burch}}, \bibinfo {author}
		{\bibfnamefont {P.~A.}\ \bibnamefont {Cassak}}, \ and\ \bibinfo {author}
		{\bibfnamefont {T.-D.}\ \bibnamefont {Phan}},\ }\href {\doibase
		10.1029/2017GL076862} {\bibfield  {journal} {\bibinfo  {journal} {Geophysical
				Research Letters}\ }\textbf {\bibinfo {volume} {45}},\ \bibinfo {pages}
		{5260} (\bibinfo {year} {2018})}\BibitemShut {NoStop}%
	\bibitem [{\citenamefont {Eastwood}\ \emph {et~al.}(2009)\citenamefont
		{Eastwood}, \citenamefont {Phan}, \citenamefont {Bale},\ and\ \citenamefont
		{Tjulin}}]{Eastwood09}%
	\BibitemOpen
	\bibfield  {author} {\bibinfo {author} {\bibfnamefont {J.~P.}\ \bibnamefont
			{Eastwood}}, \bibinfo {author} {\bibfnamefont {T.~D.}\ \bibnamefont {Phan}},
		\bibinfo {author} {\bibfnamefont {S.~D.}\ \bibnamefont {Bale}}, \ and\
		\bibinfo {author} {\bibfnamefont {A.}~\bibnamefont {Tjulin}},\ }\href
	{\doibase 10.1103/PhysRevLett.102.035001} {\bibfield  {journal} {\bibinfo
			{journal} {Phys. Rev. Lett.}\ }\textbf {\bibinfo {volume} {102}},\ \bibinfo
		{pages} {035001} (\bibinfo {year} {2009})}\BibitemShut {NoStop}%
	\bibitem [{\citenamefont {Fu}\ \emph {et~al.}(2017)\citenamefont {Fu},
		\citenamefont {Vaivads}, \citenamefont {Khotyaintsev}, \citenamefont
		{Andr\'e}, \citenamefont {Cao}, \citenamefont {Olshevsky}, \citenamefont
		{Eastwood},\ and\ \citenamefont {Retin\`o}}]{Fu17}%
	\BibitemOpen
	\bibfield  {author} {\bibinfo {author} {\bibfnamefont {H.~S.}\ \bibnamefont
			{Fu}}, \bibinfo {author} {\bibfnamefont {A.}~\bibnamefont {Vaivads}},
		\bibinfo {author} {\bibfnamefont {Y.~V.}\ \bibnamefont {Khotyaintsev}},
		\bibinfo {author} {\bibfnamefont {M.}~\bibnamefont {Andr\'e}}, \bibinfo
		{author} {\bibfnamefont {J.~B.}\ \bibnamefont {Cao}}, \bibinfo {author}
		{\bibfnamefont {V.}~\bibnamefont {Olshevsky}}, \bibinfo {author}
		{\bibfnamefont {J.~P.}\ \bibnamefont {Eastwood}}, \ and\ \bibinfo {author}
		{\bibfnamefont {A.}~\bibnamefont {Retin\`o}},\ }\href
	{https://agupubs.onlinelibrary.wiley.com/doi/abs/10.1002/2016GL071787}
	{\bibfield  {journal} {\bibinfo  {journal} {Geophysical Research Letters}\
		}\textbf {\bibinfo {volume} {44}},\ \bibinfo {pages} {37} (\bibinfo {year}
		{2017})}\BibitemShut {NoStop}%
	\bibitem [{\citenamefont {Graham}\ \emph {et~al.}(2017)\citenamefont {Graham},
		\citenamefont {Khotyaintsev}, \citenamefont {Norgren}, \citenamefont
		{Vaivads}, \citenamefont {André}, \citenamefont {Toledo-Redondo},
		\citenamefont {Lindqvist}, \citenamefont {Marklund}, \citenamefont {Ergun},
		\citenamefont {Paterson}, \citenamefont {Gershman}, \citenamefont {Giles},
		\citenamefont {Pollock}, \citenamefont {Dorelli}, \citenamefont {Avanov},
		\citenamefont {Lavraud}, \citenamefont {Saito}, \citenamefont {Magnes},
		\citenamefont {Russell}, \citenamefont {Strangeway}, \citenamefont
		{Torbert},\ and\ \citenamefont {Burch}}]{Graham17}%
	\BibitemOpen
	\bibfield  {author} {\bibinfo {author} {\bibfnamefont {D.~B.}\ \bibnamefont
			{Graham}}, \bibinfo {author} {\bibfnamefont {Y.~V.}\ \bibnamefont
			{Khotyaintsev}}, \bibinfo {author} {\bibfnamefont {C.}~\bibnamefont
			{Norgren}}, \bibinfo {author} {\bibfnamefont {A.}~\bibnamefont {Vaivads}},
		\bibinfo {author} {\bibfnamefont {M.}~\bibnamefont {André}}, \bibinfo
		{author} {\bibfnamefont {S.}~\bibnamefont {Toledo-Redondo}}, \bibinfo
		{author} {\bibfnamefont {P.-A.}\ \bibnamefont {Lindqvist}}, \bibinfo {author}
		{\bibfnamefont {G.~T.}\ \bibnamefont {Marklund}}, \bibinfo {author}
		{\bibfnamefont {R.~E.}\ \bibnamefont {Ergun}}, \bibinfo {author}
		{\bibfnamefont {W.~R.}\ \bibnamefont {Paterson}}, \bibinfo {author}
		{\bibfnamefont {D.~J.}\ \bibnamefont {Gershman}}, \bibinfo {author}
		{\bibfnamefont {B.~L.}\ \bibnamefont {Giles}}, \bibinfo {author}
		{\bibfnamefont {C.~J.}\ \bibnamefont {Pollock}}, \bibinfo {author}
		{\bibfnamefont {J.~C.}\ \bibnamefont {Dorelli}}, \bibinfo {author}
		{\bibfnamefont {L.~A.}\ \bibnamefont {Avanov}}, \bibinfo {author}
		{\bibfnamefont {B.}~\bibnamefont {Lavraud}}, \bibinfo {author} {\bibfnamefont
			{Y.}~\bibnamefont {Saito}}, \bibinfo {author} {\bibfnamefont
			{W.}~\bibnamefont {Magnes}}, \bibinfo {author} {\bibfnamefont {C.~T.}\
			\bibnamefont {Russell}}, \bibinfo {author} {\bibfnamefont {R.~J.}\
			\bibnamefont {Strangeway}}, \bibinfo {author} {\bibfnamefont {R.~B.}\
			\bibnamefont {Torbert}}, \ and\ \bibinfo {author} {\bibfnamefont {J.~L.}\
			\bibnamefont {Burch}},\ }\href {\doibase 10.1002/2016JA023572} {\bibfield
		{journal} {\bibinfo  {journal} {Journal of Geophysical Research: Space
				Physics}\ }\textbf {\bibinfo {volume} {122}},\ \bibinfo {pages} {517}
		(\bibinfo {year} {2017})}\BibitemShut {NoStop}%
	\bibitem [{\citenamefont {Ergun}\ \emph {et~al.}(2017)\citenamefont {Ergun},
		\citenamefont {Chen}, \citenamefont {Wilder}, \citenamefont {Ahmadi},
		\citenamefont {Eriksson}, \citenamefont {Usanova}, \citenamefont {Goodrich},
		\citenamefont {Holmes}, \citenamefont {Sturner}, \citenamefont {Malaspina},
		\citenamefont {Newman}, \citenamefont {Torbert}, \citenamefont {Argall},
		\citenamefont {Lindqvist}, \citenamefont {Burch}, \citenamefont {Webster},
		\citenamefont {Drake}, \citenamefont {Price}, \citenamefont {Cassak},
		\citenamefont {Swisdak}, \citenamefont {Shay}, \citenamefont {Graham},
		\citenamefont {Strangeway}, \citenamefont {Russell}, \citenamefont {Giles},
		\citenamefont {Dorelli}, \citenamefont {Gershman}, \citenamefont {Avanov},
		\citenamefont {Hesse}, \citenamefont {Lavraud}, \citenamefont {Le~Contel},
		\citenamefont {Retino}, \citenamefont {Phan}, \citenamefont {Goldman},
		\citenamefont {Stawarz}, \citenamefont {Schwartz}, \citenamefont {Eastwood},
		\citenamefont {Hwang}, \citenamefont {Nakamura},\ and\ \citenamefont
		{Wang}}]{Ergun17}%
	\BibitemOpen
	\bibfield  {author} {\bibinfo {author} {\bibfnamefont {R.~E.}\ \bibnamefont
			{Ergun}}, \bibinfo {author} {\bibfnamefont {L.-J.}\ \bibnamefont {Chen}},
		\bibinfo {author} {\bibfnamefont {F.~D.}\ \bibnamefont {Wilder}}, \bibinfo
		{author} {\bibfnamefont {N.}~\bibnamefont {Ahmadi}}, \bibinfo {author}
		{\bibfnamefont {S.}~\bibnamefont {Eriksson}}, \bibinfo {author}
		{\bibfnamefont {M.~E.}\ \bibnamefont {Usanova}}, \bibinfo {author}
		{\bibfnamefont {K.~A.}\ \bibnamefont {Goodrich}}, \bibinfo {author}
		{\bibfnamefont {J.~C.}\ \bibnamefont {Holmes}}, \bibinfo {author}
		{\bibfnamefont {A.~P.}\ \bibnamefont {Sturner}}, \bibinfo {author}
		{\bibfnamefont {D.~M.}\ \bibnamefont {Malaspina}}, \bibinfo {author}
		{\bibfnamefont {D.~L.}\ \bibnamefont {Newman}}, \bibinfo {author}
		{\bibfnamefont {R.~B.}\ \bibnamefont {Torbert}}, \bibinfo {author}
		{\bibfnamefont {M.~R.}\ \bibnamefont {Argall}}, \bibinfo {author}
		{\bibfnamefont {P.-A.}\ \bibnamefont {Lindqvist}}, \bibinfo {author}
		{\bibfnamefont {J.~L.}\ \bibnamefont {Burch}}, \bibinfo {author}
		{\bibfnamefont {J.~M.}\ \bibnamefont {Webster}}, \bibinfo {author}
		{\bibfnamefont {J.~F.}\ \bibnamefont {Drake}}, \bibinfo {author}
		{\bibfnamefont {L.}~\bibnamefont {Price}}, \bibinfo {author} {\bibfnamefont
			{P.~A.}\ \bibnamefont {Cassak}}, \bibinfo {author} {\bibfnamefont
			{M.}~\bibnamefont {Swisdak}}, \bibinfo {author} {\bibfnamefont {M.~A.}\
			\bibnamefont {Shay}}, \bibinfo {author} {\bibfnamefont {D.~B.}\ \bibnamefont
			{Graham}}, \bibinfo {author} {\bibfnamefont {R.~J.}\ \bibnamefont
			{Strangeway}}, \bibinfo {author} {\bibfnamefont {C.~T.}\ \bibnamefont
			{Russell}}, \bibinfo {author} {\bibfnamefont {B.~L.}\ \bibnamefont {Giles}},
		\bibinfo {author} {\bibfnamefont {J.~C.}\ \bibnamefont {Dorelli}}, \bibinfo
		{author} {\bibfnamefont {D.}~\bibnamefont {Gershman}}, \bibinfo {author}
		{\bibfnamefont {L.}~\bibnamefont {Avanov}}, \bibinfo {author} {\bibfnamefont
			{M.}~\bibnamefont {Hesse}}, \bibinfo {author} {\bibfnamefont
			{B.}~\bibnamefont {Lavraud}}, \bibinfo {author} {\bibfnamefont
			{O.}~\bibnamefont {Le~Contel}}, \bibinfo {author} {\bibfnamefont
			{A.}~\bibnamefont {Retino}}, \bibinfo {author} {\bibfnamefont {T.~D.}\
			\bibnamefont {Phan}}, \bibinfo {author} {\bibfnamefont {M.~V.}\ \bibnamefont
			{Goldman}}, \bibinfo {author} {\bibfnamefont {J.~E.}\ \bibnamefont
			{Stawarz}}, \bibinfo {author} {\bibfnamefont {S.~J.}\ \bibnamefont
			{Schwartz}}, \bibinfo {author} {\bibfnamefont {J.~P.}\ \bibnamefont
			{Eastwood}}, \bibinfo {author} {\bibfnamefont {K.-J.}\ \bibnamefont {Hwang}},
		\bibinfo {author} {\bibfnamefont {R.}~\bibnamefont {Nakamura}}, \ and\
		\bibinfo {author} {\bibfnamefont {S.}~\bibnamefont {Wang}},\ }\href {\doibase
		10.1002/2016GL072493} {\bibfield  {journal} {\bibinfo  {journal} {Geophysical
				Research Letters}\ }\textbf {\bibinfo {volume} {44}},\ \bibinfo {pages}
		{2978} (\bibinfo {year} {2017})}\BibitemShut {NoStop}%
	\bibitem [{\citenamefont {Osman}\ \emph {et~al.}(2015)\citenamefont {Osman},
		\citenamefont {Kiyani}, \citenamefont {Matthaeus}, \citenamefont {Hnat},
		\citenamefont {Chapman},\ and\ \citenamefont {Khotyaintsev}}]{Osman15}%
	\BibitemOpen
	\bibfield  {author} {\bibinfo {author} {\bibfnamefont {K.~T.}\ \bibnamefont
			{Osman}}, \bibinfo {author} {\bibfnamefont {K.~H.}\ \bibnamefont {Kiyani}},
		\bibinfo {author} {\bibfnamefont {W.~H.}\ \bibnamefont {Matthaeus}}, \bibinfo
		{author} {\bibfnamefont {B.}~\bibnamefont {Hnat}}, \bibinfo {author}
		{\bibfnamefont {S.~C.}\ \bibnamefont {Chapman}}, \ and\ \bibinfo {author}
		{\bibfnamefont {Y.~V.}\ \bibnamefont {Khotyaintsev}},\ }\href
	{http://stacks.iop.org/2041-8205/815/i=2/a=L24} {\bibfield  {journal}
		{\bibinfo  {journal} {The Astrophysical Journal Letters}\ }\textbf {\bibinfo
			{volume} {815}},\ \bibinfo {pages} {L24} (\bibinfo {year}
		{2015})}\BibitemShut {NoStop}%
	\bibitem [{\citenamefont {Phan}\ \emph {et~al.}(2016)\citenamefont {Phan},
		\citenamefont {Eastwood}, \citenamefont {Cassak}, \citenamefont {Øieroset},
		\citenamefont {Gosling}, \citenamefont {Gershman}, \citenamefont {Mozer},
		\citenamefont {Shay}, \citenamefont {Fujimoto}, \citenamefont {Daughton},
		\citenamefont {Drake}, \citenamefont {Burch}, \citenamefont {Torbert},
		\citenamefont {Ergun}, \citenamefont {Chen}, \citenamefont {Wang},
		\citenamefont {Pollock}, \citenamefont {Dorelli}, \citenamefont {Lavraud},
		\citenamefont {Giles}, \citenamefont {Moore}, \citenamefont {Saito},
		\citenamefont {Avanov}, \citenamefont {Paterson}, \citenamefont {Strangeway},
		\citenamefont {Russell}, \citenamefont {Khotyaintsev}, \citenamefont
		{Lindqvist}, \citenamefont {Oka},\ and\ \citenamefont {Wilder}}]{Phan16}%
	\BibitemOpen
	\bibfield  {author} {\bibinfo {author} {\bibfnamefont {T.~D.}\ \bibnamefont
			{Phan}}, \bibinfo {author} {\bibfnamefont {J.~P.}\ \bibnamefont {Eastwood}},
		\bibinfo {author} {\bibfnamefont {P.~A.}\ \bibnamefont {Cassak}}, \bibinfo
		{author} {\bibfnamefont {M.}~\bibnamefont {Øieroset}}, \bibinfo {author}
		{\bibfnamefont {J.~T.}\ \bibnamefont {Gosling}}, \bibinfo {author}
		{\bibfnamefont {D.~J.}\ \bibnamefont {Gershman}}, \bibinfo {author}
		{\bibfnamefont {F.~S.}\ \bibnamefont {Mozer}}, \bibinfo {author}
		{\bibfnamefont {M.~A.}\ \bibnamefont {Shay}}, \bibinfo {author}
		{\bibfnamefont {M.}~\bibnamefont {Fujimoto}}, \bibinfo {author}
		{\bibfnamefont {W.}~\bibnamefont {Daughton}}, \bibinfo {author}
		{\bibfnamefont {J.~F.}\ \bibnamefont {Drake}}, \bibinfo {author}
		{\bibfnamefont {J.~L.}\ \bibnamefont {Burch}}, \bibinfo {author}
		{\bibfnamefont {R.~B.}\ \bibnamefont {Torbert}}, \bibinfo {author}
		{\bibfnamefont {R.~E.}\ \bibnamefont {Ergun}}, \bibinfo {author}
		{\bibfnamefont {L.~J.}\ \bibnamefont {Chen}}, \bibinfo {author}
		{\bibfnamefont {S.}~\bibnamefont {Wang}}, \bibinfo {author} {\bibfnamefont
			{C.}~\bibnamefont {Pollock}}, \bibinfo {author} {\bibfnamefont {J.~C.}\
			\bibnamefont {Dorelli}}, \bibinfo {author} {\bibfnamefont {B.}~\bibnamefont
			{Lavraud}}, \bibinfo {author} {\bibfnamefont {B.~L.}\ \bibnamefont {Giles}},
		\bibinfo {author} {\bibfnamefont {T.~E.}\ \bibnamefont {Moore}}, \bibinfo
		{author} {\bibfnamefont {Y.}~\bibnamefont {Saito}}, \bibinfo {author}
		{\bibfnamefont {L.~A.}\ \bibnamefont {Avanov}}, \bibinfo {author}
		{\bibfnamefont {W.}~\bibnamefont {Paterson}}, \bibinfo {author}
		{\bibfnamefont {R.~J.}\ \bibnamefont {Strangeway}}, \bibinfo {author}
		{\bibfnamefont {C.~T.}\ \bibnamefont {Russell}}, \bibinfo {author}
		{\bibfnamefont {Y.}~\bibnamefont {Khotyaintsev}}, \bibinfo {author}
		{\bibfnamefont {P.~A.}\ \bibnamefont {Lindqvist}}, \bibinfo {author}
		{\bibfnamefont {M.}~\bibnamefont {Oka}}, \ and\ \bibinfo {author}
		{\bibfnamefont {F.~D.}\ \bibnamefont {Wilder}},\ }\href {\doibase
		10.1002/2016GL069212} {\bibfield  {journal} {\bibinfo  {journal} {Geophysical
				Research Letters}\ }\textbf {\bibinfo {volume} {43}},\ \bibinfo {pages}
		{6060} (\bibinfo {year} {2016})}\BibitemShut {NoStop}%
	\bibitem [{\citenamefont {Burch}\ \emph {et~al.}(2018)\citenamefont {Burch},
		\citenamefont {Ergun}, \citenamefont {Cassak}, \citenamefont {Webster},
		\citenamefont {Torbert}, \citenamefont {Giles}, \citenamefont {Dorelli},
		\citenamefont {Rager}, \citenamefont {Hwang}, \citenamefont {Phan},
		\citenamefont {Genestreti}, \citenamefont {Allen}, \citenamefont {Chen},
		\citenamefont {Wang}, \citenamefont {Gershman}, \citenamefont {Le~Contel},
		\citenamefont {Russell}, \citenamefont {Strangeway}, \citenamefont {Wilder},
		\citenamefont {Graham}, \citenamefont {Hesse}, \citenamefont {Drake},
		\citenamefont {Swisdak}, \citenamefont {Price}, \citenamefont {Shay},
		\citenamefont {Lindqvist}, \citenamefont {Pollock}, \citenamefont {Denton},\
		and\ \citenamefont {Newman}}]{Burch18}%
	\BibitemOpen
	\bibfield  {author} {\bibinfo {author} {\bibfnamefont {J.~L.}\ \bibnamefont
			{Burch}}, \bibinfo {author} {\bibfnamefont {R.~E.}\ \bibnamefont {Ergun}},
		\bibinfo {author} {\bibfnamefont {P.~A.}\ \bibnamefont {Cassak}}, \bibinfo
		{author} {\bibfnamefont {J.~M.}\ \bibnamefont {Webster}}, \bibinfo {author}
		{\bibfnamefont {R.~B.}\ \bibnamefont {Torbert}}, \bibinfo {author}
		{\bibfnamefont {B.~L.}\ \bibnamefont {Giles}}, \bibinfo {author}
		{\bibfnamefont {J.~C.}\ \bibnamefont {Dorelli}}, \bibinfo {author}
		{\bibfnamefont {A.~C.}\ \bibnamefont {Rager}}, \bibinfo {author}
		{\bibfnamefont {K.-J.}\ \bibnamefont {Hwang}}, \bibinfo {author}
		{\bibfnamefont {T.~D.}\ \bibnamefont {Phan}}, \bibinfo {author}
		{\bibfnamefont {K.~J.}\ \bibnamefont {Genestreti}}, \bibinfo {author}
		{\bibfnamefont {R.~C.}\ \bibnamefont {Allen}}, \bibinfo {author}
		{\bibfnamefont {L.-J.}\ \bibnamefont {Chen}}, \bibinfo {author}
		{\bibfnamefont {S.}~\bibnamefont {Wang}}, \bibinfo {author} {\bibfnamefont
			{D.}~\bibnamefont {Gershman}}, \bibinfo {author} {\bibfnamefont
			{O.}~\bibnamefont {Le~Contel}}, \bibinfo {author} {\bibfnamefont {C.~T.}\
			\bibnamefont {Russell}}, \bibinfo {author} {\bibfnamefont {R.~J.}\
			\bibnamefont {Strangeway}}, \bibinfo {author} {\bibfnamefont {F.~D.}\
			\bibnamefont {Wilder}}, \bibinfo {author} {\bibfnamefont {D.~B.}\
			\bibnamefont {Graham}}, \bibinfo {author} {\bibfnamefont {M.}~\bibnamefont
			{Hesse}}, \bibinfo {author} {\bibfnamefont {J.~F.}\ \bibnamefont {Drake}},
		\bibinfo {author} {\bibfnamefont {M.}~\bibnamefont {Swisdak}}, \bibinfo
		{author} {\bibfnamefont {L.~M.}\ \bibnamefont {Price}}, \bibinfo {author}
		{\bibfnamefont {M.~A.}\ \bibnamefont {Shay}}, \bibinfo {author}
		{\bibfnamefont {P.-A.}\ \bibnamefont {Lindqvist}}, \bibinfo {author}
		{\bibfnamefont {C.~J.}\ \bibnamefont {Pollock}}, \bibinfo {author}
		{\bibfnamefont {R.~E.}\ \bibnamefont {Denton}}, \ and\ \bibinfo {author}
		{\bibfnamefont {D.~L.}\ \bibnamefont {Newman}},\ }\href {\doibase
		10.1002/2017GL076809} {\bibfield  {journal} {\bibinfo  {journal} {Geophysical
				Research Letters}\ }\textbf {\bibinfo {volume} {45}},\ \bibinfo {pages}
		{1237} (\bibinfo {year} {2018})}\BibitemShut {NoStop}%
	\bibitem [{\citenamefont {Acu{\~{n}}a}\ \emph {et~al.}(1995)\citenamefont
		{Acu{\~{n}}a}, \citenamefont {Ogilvie}, \citenamefont {Baker}, \citenamefont
		{Curtis}, \citenamefont {Fairfield},\ and\ \citenamefont {Mish}}]{Acuna95}%
	\BibitemOpen
	\bibfield  {author} {\bibinfo {author} {\bibfnamefont {M.~H.}\ \bibnamefont
			{Acu{\~{n}}a}}, \bibinfo {author} {\bibfnamefont {K.~W.}\ \bibnamefont
			{Ogilvie}}, \bibinfo {author} {\bibfnamefont {D.~N.}\ \bibnamefont {Baker}},
		\bibinfo {author} {\bibfnamefont {S.~A.}\ \bibnamefont {Curtis}}, \bibinfo
		{author} {\bibfnamefont {D.~H.}\ \bibnamefont {Fairfield}}, \ and\ \bibinfo
		{author} {\bibfnamefont {W.~H.}\ \bibnamefont {Mish}},\ }\href {\doibase
		10.1007/BF00751323} {\bibfield  {journal} {\bibinfo  {journal} {Space Science
				Reviews}\ }\textbf {\bibinfo {volume} {71}},\ \bibinfo {pages} {5} (\bibinfo
		{year} {1995})}\BibitemShut {NoStop}%
	\bibitem [{\citenamefont {Russell}\ \emph {et~al.}(2016)\citenamefont
		{Russell}, \citenamefont {Anderson}, \citenamefont {Baumjohann},
		\citenamefont {Bromund}, \citenamefont {Dearborn}, \citenamefont {Fischer},
		\citenamefont {Le}, \citenamefont {Leinweber}, \citenamefont {Leneman},
		\citenamefont {Magnes}, \citenamefont {Means}, \citenamefont {Moldwin},
		\citenamefont {Nakamura}, \citenamefont {Pierce}, \citenamefont {Plaschke},
		\citenamefont {Rowe}, \citenamefont {Slavin}, \citenamefont {Strangeway},
		\citenamefont {Torbert}, \citenamefont {Hagen}, \citenamefont {Jernej},
		\citenamefont {Valavanoglou},\ and\ \citenamefont {Richter}}]{Russell2016}%
	\BibitemOpen
	\bibfield  {author} {\bibinfo {author} {\bibfnamefont {C.~T.}\ \bibnamefont
			{Russell}}, \bibinfo {author} {\bibfnamefont {B.~J.}\ \bibnamefont
			{Anderson}}, \bibinfo {author} {\bibfnamefont {W.}~\bibnamefont
			{Baumjohann}}, \bibinfo {author} {\bibfnamefont {K.~R.}\ \bibnamefont
			{Bromund}}, \bibinfo {author} {\bibfnamefont {D.}~\bibnamefont {Dearborn}},
		\bibinfo {author} {\bibfnamefont {D.}~\bibnamefont {Fischer}}, \bibinfo
		{author} {\bibfnamefont {G.}~\bibnamefont {Le}}, \bibinfo {author}
		{\bibfnamefont {H.~K.}\ \bibnamefont {Leinweber}}, \bibinfo {author}
		{\bibfnamefont {D.}~\bibnamefont {Leneman}}, \bibinfo {author} {\bibfnamefont
			{W.}~\bibnamefont {Magnes}}, \bibinfo {author} {\bibfnamefont {J.~D.}\
			\bibnamefont {Means}}, \bibinfo {author} {\bibfnamefont {M.~B.}\ \bibnamefont
			{Moldwin}}, \bibinfo {author} {\bibfnamefont {R.}~\bibnamefont {Nakamura}},
		\bibinfo {author} {\bibfnamefont {D.}~\bibnamefont {Pierce}}, \bibinfo
		{author} {\bibfnamefont {F.}~\bibnamefont {Plaschke}}, \bibinfo {author}
		{\bibfnamefont {K.~M.}\ \bibnamefont {Rowe}}, \bibinfo {author}
		{\bibfnamefont {J.~A.}\ \bibnamefont {Slavin}}, \bibinfo {author}
		{\bibfnamefont {R.~J.}\ \bibnamefont {Strangeway}}, \bibinfo {author}
		{\bibfnamefont {R.}~\bibnamefont {Torbert}}, \bibinfo {author} {\bibfnamefont
			{C.}~\bibnamefont {Hagen}}, \bibinfo {author} {\bibfnamefont
			{I.}~\bibnamefont {Jernej}}, \bibinfo {author} {\bibfnamefont
			{A.}~\bibnamefont {Valavanoglou}}, \ and\ \bibinfo {author} {\bibfnamefont
			{I.}~\bibnamefont {Richter}},\ }\href {\doibase 10.1007/s11214-014-0057-3}
	{\bibfield  {journal} {\bibinfo  {journal} {Space Science Reviews}\ }\textbf
		{\bibinfo {volume} {199}},\ \bibinfo {pages} {189} (\bibinfo {year}
		{2016})}\BibitemShut {NoStop}%
	\bibitem [{\citenamefont {Ergun}\ \emph {et~al.}(2016)\citenamefont {Ergun},
		\citenamefont {Tucker}, \citenamefont {Westfall}, \citenamefont {Goodrich},
		\citenamefont {Malaspina}, \citenamefont {Summers}, \citenamefont {Wallace},
		\citenamefont {Karlsson}, \citenamefont {Mack}, \citenamefont {Brennan},
		\citenamefont {Pyke}, \citenamefont {Withnell}, \citenamefont {Torbert},
		\citenamefont {Macri}, \citenamefont {Rau}, \citenamefont {Dors},
		\citenamefont {Needell}, \citenamefont {Lindqvist}, \citenamefont {Olsson},\
		and\ \citenamefont {Cully}}]{Ergun16}%
	\BibitemOpen
	\bibfield  {author} {\bibinfo {author} {\bibfnamefont {R.~E.}\ \bibnamefont
			{Ergun}}, \bibinfo {author} {\bibfnamefont {S.}~\bibnamefont {Tucker}},
		\bibinfo {author} {\bibfnamefont {J.}~\bibnamefont {Westfall}}, \bibinfo
		{author} {\bibfnamefont {K.~A.}\ \bibnamefont {Goodrich}}, \bibinfo {author}
		{\bibfnamefont {D.~M.}\ \bibnamefont {Malaspina}}, \bibinfo {author}
		{\bibfnamefont {D.}~\bibnamefont {Summers}}, \bibinfo {author} {\bibfnamefont
			{J.}~\bibnamefont {Wallace}}, \bibinfo {author} {\bibfnamefont
			{M.}~\bibnamefont {Karlsson}}, \bibinfo {author} {\bibfnamefont
			{J.}~\bibnamefont {Mack}}, \bibinfo {author} {\bibfnamefont {N.}~\bibnamefont
			{Brennan}}, \bibinfo {author} {\bibfnamefont {B.}~\bibnamefont {Pyke}},
		\bibinfo {author} {\bibfnamefont {P.}~\bibnamefont {Withnell}}, \bibinfo
		{author} {\bibfnamefont {R.}~\bibnamefont {Torbert}}, \bibinfo {author}
		{\bibfnamefont {J.}~\bibnamefont {Macri}}, \bibinfo {author} {\bibfnamefont
			{D.}~\bibnamefont {Rau}}, \bibinfo {author} {\bibfnamefont {I.}~\bibnamefont
			{Dors}}, \bibinfo {author} {\bibfnamefont {J.}~\bibnamefont {Needell}},
		\bibinfo {author} {\bibfnamefont {P.-A.}\ \bibnamefont {Lindqvist}}, \bibinfo
		{author} {\bibfnamefont {G.}~\bibnamefont {Olsson}}, \ and\ \bibinfo {author}
		{\bibfnamefont {C.~M.}\ \bibnamefont {Cully}},\ }\href {\doibase
		10.1007/s11214-014-0115-x} {\bibfield  {journal} {\bibinfo  {journal} {Space
				Science Reviews}\ }\textbf {\bibinfo {volume} {199}},\ \bibinfo {pages} {167}
		(\bibinfo {year} {2016})}\BibitemShut {NoStop}%
	\bibitem [{\citenamefont {Lindqvist}\ \emph {et~al.}(2016)\citenamefont
		{Lindqvist}, \citenamefont {Olsson}, \citenamefont {Torbert}, \citenamefont
		{King}, \citenamefont {Granoff}, \citenamefont {Rau}, \citenamefont
		{Needell}, \citenamefont {Turco}, \citenamefont {Dors}, \citenamefont
		{Beckman}, \citenamefont {Macri}, \citenamefont {Frost}, \citenamefont
		{Salwen}, \citenamefont {Eriksson}, \citenamefont {{\AA}hl{\'e}n},
		\citenamefont {Khotyaintsev}, \citenamefont {Porter}, \citenamefont
		{Lappalainen}, \citenamefont {Ergun}, \citenamefont {Wermeer},\ and\
		\citenamefont {Tucker}}]{Lindqvist16}%
	\BibitemOpen
	\bibfield  {author} {\bibinfo {author} {\bibfnamefont {P.-A.}\ \bibnamefont
			{Lindqvist}}, \bibinfo {author} {\bibfnamefont {G.}~\bibnamefont {Olsson}},
		\bibinfo {author} {\bibfnamefont {R.~B.}\ \bibnamefont {Torbert}}, \bibinfo
		{author} {\bibfnamefont {B.}~\bibnamefont {King}}, \bibinfo {author}
		{\bibfnamefont {M.}~\bibnamefont {Granoff}}, \bibinfo {author} {\bibfnamefont
			{D.}~\bibnamefont {Rau}}, \bibinfo {author} {\bibfnamefont {G.}~\bibnamefont
			{Needell}}, \bibinfo {author} {\bibfnamefont {S.}~\bibnamefont {Turco}},
		\bibinfo {author} {\bibfnamefont {I.}~\bibnamefont {Dors}}, \bibinfo {author}
		{\bibfnamefont {P.}~\bibnamefont {Beckman}}, \bibinfo {author} {\bibfnamefont
			{J.}~\bibnamefont {Macri}}, \bibinfo {author} {\bibfnamefont
			{C.}~\bibnamefont {Frost}}, \bibinfo {author} {\bibfnamefont
			{J.}~\bibnamefont {Salwen}}, \bibinfo {author} {\bibfnamefont
			{A.}~\bibnamefont {Eriksson}}, \bibinfo {author} {\bibfnamefont
			{L.}~\bibnamefont {{\AA}hl{\'e}n}}, \bibinfo {author} {\bibfnamefont {Y.~V.}\
			\bibnamefont {Khotyaintsev}}, \bibinfo {author} {\bibfnamefont
			{J.}~\bibnamefont {Porter}}, \bibinfo {author} {\bibfnamefont
			{K.}~\bibnamefont {Lappalainen}}, \bibinfo {author} {\bibfnamefont {R.~E.}\
			\bibnamefont {Ergun}}, \bibinfo {author} {\bibfnamefont {W.}~\bibnamefont
			{Wermeer}}, \ and\ \bibinfo {author} {\bibfnamefont {S.}~\bibnamefont
			{Tucker}},\ }\href {\doibase 10.1007/s11214-014-0116-9} {\bibfield  {journal}
		{\bibinfo  {journal} {Space Science Reviews}\ }\textbf {\bibinfo {volume}
			{199}},\ \bibinfo {pages} {137} (\bibinfo {year} {2016})}\BibitemShut
	{NoStop}%
	\bibitem [{\citenamefont {Pollock}\ \emph {et~al.}(2016)\citenamefont
		{Pollock}, \citenamefont {Moore}, \citenamefont {Jacques}, \citenamefont
		{Burch}, \citenamefont {Gliese}, \citenamefont {Saito}, \citenamefont
		{Omoto}, \citenamefont {Avanov}, \citenamefont {Barrie}, \citenamefont
		{Coffey}, \citenamefont {Dorelli}, \citenamefont {Gershman}, \citenamefont
		{Giles}, \citenamefont {Rosnack}, \citenamefont {Salo}, \citenamefont
		{Yokota}, \citenamefont {Adrian}, \citenamefont {Aoustin}, \citenamefont
		{Auletti}, \citenamefont {Aung}, \citenamefont {Bigio}, \citenamefont {Cao},
		\citenamefont {Chandler}, \citenamefont {Chornay}, \citenamefont {Christian},
		\citenamefont {Clark}, \citenamefont {Collinson}, \citenamefont {Corris},
		\citenamefont {De Los Santos}, \citenamefont {Devlin}, \citenamefont
		{Diaz}, \citenamefont {Dickerson}, \citenamefont {Dickson}, \citenamefont
		{Diekmann}, \citenamefont {Diggs}, \citenamefont {Duncan}, \citenamefont
		{Figueroa-Vinas}, \citenamefont {Firman}, \citenamefont {Freeman},
		\citenamefont {Galassi}, \citenamefont {Garcia}, \citenamefont {Goodhart},
		\citenamefont {Guererro}, \citenamefont {Hageman}, \citenamefont {Hanley},
		\citenamefont {Hemminger}, \citenamefont {Holland}, \citenamefont {Hutchins},
		\citenamefont {James}, \citenamefont {Jones}, \citenamefont {Kreisler},
		\citenamefont {Kujawski}, \citenamefont {Lavu}, \citenamefont {Lobell},
		\citenamefont {LeCompte}, \citenamefont {Lukemire}, \citenamefont
		{MacDonald}, \citenamefont {Mariano}, \citenamefont {Mukai}, \citenamefont
		{Narayanan}, \citenamefont {Nguyan}, \citenamefont {Onizuka}, \citenamefont
		{Paterson}, \citenamefont {Persyn}, \citenamefont {Piepgrass}, \citenamefont
		{Cheney}, \citenamefont {Rager}, \citenamefont {Raghuram}, \citenamefont
		{Ramil}, \citenamefont {Reichenthal}, \citenamefont {Rodriguez},
		\citenamefont {Rouzaud}, \citenamefont {Rucker}, \citenamefont {Saito},
		\citenamefont {Samara}, \citenamefont {Sauvaud}, \citenamefont {Schuster},
		\citenamefont {Shappirio}, \citenamefont {Shelton}, \citenamefont {Sher},
		\citenamefont {Smith}, \citenamefont {Smith}, \citenamefont {Smith},
		\citenamefont {Steinfeld}, \citenamefont {Szymkiewicz}, \citenamefont
		{Tanimoto}, \citenamefont {Taylor}, \citenamefont {Tucker}, \citenamefont
		{Tull}, \citenamefont {Uhl}, \citenamefont {Vloet}, \citenamefont {Walpole},
		\citenamefont {Weidner}, \citenamefont {White}, \citenamefont {Winkert},
		\citenamefont {Yeh},\ and\ \citenamefont {Zeuch}}]{Pollock16}%
	\BibitemOpen
	\bibfield  {author} {\bibinfo {author} {\bibfnamefont {C.}~\bibnamefont
			{Pollock}}, \bibinfo {author} {\bibfnamefont {T.}~\bibnamefont {Moore}},
		\bibinfo {author} {\bibfnamefont {A.}~\bibnamefont {Jacques}}, \bibinfo
		{author} {\bibfnamefont {J.}~\bibnamefont {Burch}}, \bibinfo {author}
		{\bibfnamefont {U.}~\bibnamefont {Gliese}}, \bibinfo {author} {\bibfnamefont
			{Y.}~\bibnamefont {Saito}}, \bibinfo {author} {\bibfnamefont
			{T.}~\bibnamefont {Omoto}}, \bibinfo {author} {\bibfnamefont
			{L.}~\bibnamefont {Avanov}}, \bibinfo {author} {\bibfnamefont
			{A.}~\bibnamefont {Barrie}}, \bibinfo {author} {\bibfnamefont
			{V.}~\bibnamefont {Coffey}}, \bibinfo {author} {\bibfnamefont
			{J.}~\bibnamefont {Dorelli}}, \bibinfo {author} {\bibfnamefont
			{D.}~\bibnamefont {Gershman}}, \bibinfo {author} {\bibfnamefont
			{B.}~\bibnamefont {Giles}}, \bibinfo {author} {\bibfnamefont
			{T.}~\bibnamefont {Rosnack}}, \bibinfo {author} {\bibfnamefont
			{C.}~\bibnamefont {Salo}}, \bibinfo {author} {\bibfnamefont {S.}~\bibnamefont
			{Yokota}}, \bibinfo {author} {\bibfnamefont {M.}~\bibnamefont {Adrian}},
		\bibinfo {author} {\bibfnamefont {C.}~\bibnamefont {Aoustin}}, \bibinfo
		{author} {\bibfnamefont {C.}~\bibnamefont {Auletti}}, \bibinfo {author}
		{\bibfnamefont {S.}~\bibnamefont {Aung}}, \bibinfo {author} {\bibfnamefont
			{V.}~\bibnamefont {Bigio}}, \bibinfo {author} {\bibfnamefont
			{N.}~\bibnamefont {Cao}}, \bibinfo {author} {\bibfnamefont {M.}~\bibnamefont
			{Chandler}}, \bibinfo {author} {\bibfnamefont {D.}~\bibnamefont {Chornay}},
		\bibinfo {author} {\bibfnamefont {K.}~\bibnamefont {Christian}}, \bibinfo
		{author} {\bibfnamefont {G.}~\bibnamefont {Clark}}, \bibinfo {author}
		{\bibfnamefont {G.}~\bibnamefont {Collinson}}, \bibinfo {author}
		{\bibfnamefont {T.}~\bibnamefont {Corris}}, \bibinfo {author} {\bibfnamefont
			{A.}~\bibnamefont {De Los Santos}}, \bibinfo {author} {\bibfnamefont
			{R.}~\bibnamefont {Devlin}}, \bibinfo {author} {\bibfnamefont
			{T.}~\bibnamefont {Diaz}}, \bibinfo {author} {\bibfnamefont {T.}~\bibnamefont
			{Dickerson}}, \bibinfo {author} {\bibfnamefont {C.}~\bibnamefont {Dickson}},
		\bibinfo {author} {\bibfnamefont {A.}~\bibnamefont {Diekmann}}, \bibinfo
		{author} {\bibfnamefont {F.}~\bibnamefont {Diggs}}, \bibinfo {author}
		{\bibfnamefont {C.}~\bibnamefont {Duncan}}, \bibinfo {author} {\bibfnamefont
			{A.}~\bibnamefont {Figueroa-Vinas}}, \bibinfo {author} {\bibfnamefont
			{C.}~\bibnamefont {Firman}}, \bibinfo {author} {\bibfnamefont
			{M.}~\bibnamefont {Freeman}}, \bibinfo {author} {\bibfnamefont
			{N.}~\bibnamefont {Galassi}}, \bibinfo {author} {\bibfnamefont
			{K.}~\bibnamefont {Garcia}}, \bibinfo {author} {\bibfnamefont
			{G.}~\bibnamefont {Goodhart}}, \bibinfo {author} {\bibfnamefont
			{D.}~\bibnamefont {Guererro}}, \bibinfo {author} {\bibfnamefont
			{J.}~\bibnamefont {Hageman}}, \bibinfo {author} {\bibfnamefont
			{J.}~\bibnamefont {Hanley}}, \bibinfo {author} {\bibfnamefont
			{E.}~\bibnamefont {Hemminger}}, \bibinfo {author} {\bibfnamefont
			{M.}~\bibnamefont {Holland}}, \bibinfo {author} {\bibfnamefont
			{M.}~\bibnamefont {Hutchins}}, \bibinfo {author} {\bibfnamefont
			{T.}~\bibnamefont {James}}, \bibinfo {author} {\bibfnamefont
			{W.}~\bibnamefont {Jones}}, \bibinfo {author} {\bibfnamefont
			{S.}~\bibnamefont {Kreisler}}, \bibinfo {author} {\bibfnamefont
			{J.}~\bibnamefont {Kujawski}}, \bibinfo {author} {\bibfnamefont
			{V.}~\bibnamefont {Lavu}}, \bibinfo {author} {\bibfnamefont {J.}~\bibnamefont
			{Lobell}}, \bibinfo {author} {\bibfnamefont {E.}~\bibnamefont {LeCompte}},
		\bibinfo {author} {\bibfnamefont {A.}~\bibnamefont {Lukemire}}, \bibinfo
		{author} {\bibfnamefont {E.}~\bibnamefont {MacDonald}}, \bibinfo {author}
		{\bibfnamefont {A.}~\bibnamefont {Mariano}}, \bibinfo {author} {\bibfnamefont
			{T.}~\bibnamefont {Mukai}}, \bibinfo {author} {\bibfnamefont
			{K.}~\bibnamefont {Narayanan}}, \bibinfo {author} {\bibfnamefont
			{Q.}~\bibnamefont {Nguyan}}, \bibinfo {author} {\bibfnamefont
			{M.}~\bibnamefont {Onizuka}}, \bibinfo {author} {\bibfnamefont
			{W.}~\bibnamefont {Paterson}}, \bibinfo {author} {\bibfnamefont
			{S.}~\bibnamefont {Persyn}}, \bibinfo {author} {\bibfnamefont
			{B.}~\bibnamefont {Piepgrass}}, \bibinfo {author} {\bibfnamefont
			{F.}~\bibnamefont {Cheney}}, \bibinfo {author} {\bibfnamefont
			{A.}~\bibnamefont {Rager}}, \bibinfo {author} {\bibfnamefont
			{T.}~\bibnamefont {Raghuram}}, \bibinfo {author} {\bibfnamefont
			{A.}~\bibnamefont {Ramil}}, \bibinfo {author} {\bibfnamefont
			{L.}~\bibnamefont {Reichenthal}}, \bibinfo {author} {\bibfnamefont
			{H.}~\bibnamefont {Rodriguez}}, \bibinfo {author} {\bibfnamefont
			{J.}~\bibnamefont {Rouzaud}}, \bibinfo {author} {\bibfnamefont
			{A.}~\bibnamefont {Rucker}}, \bibinfo {author} {\bibfnamefont
			{Y.}~\bibnamefont {Saito}}, \bibinfo {author} {\bibfnamefont
			{M.}~\bibnamefont {Samara}}, \bibinfo {author} {\bibfnamefont {J.-A.}\
			\bibnamefont {Sauvaud}}, \bibinfo {author} {\bibfnamefont {D.}~\bibnamefont
			{Schuster}}, \bibinfo {author} {\bibfnamefont {M.}~\bibnamefont {Shappirio}},
		\bibinfo {author} {\bibfnamefont {K.}~\bibnamefont {Shelton}}, \bibinfo
		{author} {\bibfnamefont {D.}~\bibnamefont {Sher}}, \bibinfo {author}
		{\bibfnamefont {D.}~\bibnamefont {Smith}}, \bibinfo {author} {\bibfnamefont
			{K.}~\bibnamefont {Smith}}, \bibinfo {author} {\bibfnamefont
			{S.}~\bibnamefont {Smith}}, \bibinfo {author} {\bibfnamefont
			{D.}~\bibnamefont {Steinfeld}}, \bibinfo {author} {\bibfnamefont
			{R.}~\bibnamefont {Szymkiewicz}}, \bibinfo {author} {\bibfnamefont
			{K.}~\bibnamefont {Tanimoto}}, \bibinfo {author} {\bibfnamefont
			{J.}~\bibnamefont {Taylor}}, \bibinfo {author} {\bibfnamefont
			{C.}~\bibnamefont {Tucker}}, \bibinfo {author} {\bibfnamefont
			{K.}~\bibnamefont {Tull}}, \bibinfo {author} {\bibfnamefont {A.}~\bibnamefont
			{Uhl}}, \bibinfo {author} {\bibfnamefont {J.}~\bibnamefont {Vloet}}, \bibinfo
		{author} {\bibfnamefont {P.}~\bibnamefont {Walpole}}, \bibinfo {author}
		{\bibfnamefont {S.}~\bibnamefont {Weidner}}, \bibinfo {author} {\bibfnamefont
			{D.}~\bibnamefont {White}}, \bibinfo {author} {\bibfnamefont
			{G.}~\bibnamefont {Winkert}}, \bibinfo {author} {\bibfnamefont {P.-S.}\
			\bibnamefont {Yeh}}, \ and\ \bibinfo {author} {\bibfnamefont
			{M.}~\bibnamefont {Zeuch}},\ }\href {\doibase 10.1007/s11214-016-0245-4}
	{\bibfield  {journal} {\bibinfo  {journal} {Space Science Reviews}\ }\textbf
		{\bibinfo {volume} {199}},\ \bibinfo {pages} {331} (\bibinfo {year}
		{2016})}\BibitemShut {NoStop}%
	\bibitem [{\citenamefont {Eastman}\ and\ \citenamefont
		{Hones~Jr.}(1979)}]{Eastman79}%
	\BibitemOpen
	\bibfield  {author} {\bibinfo {author} {\bibfnamefont {T.~E.}\ \bibnamefont
			{Eastman}}\ and\ \bibinfo {author} {\bibfnamefont {E.~W.}\ \bibnamefont
			{Hones~Jr.}},\ }\href {\doibase 10.1029/JA084iA05p02019} {\bibfield
		{journal} {\bibinfo  {journal} {Journal of Geophysical Research: Space
				Physics}\ }\textbf {\bibinfo {volume} {84}},\ \bibinfo {pages} {2019}
		(\bibinfo {year} {1979})}\BibitemShut {NoStop}%
	\bibitem [{\citenamefont {Paschmann}\ and\ \citenamefont
		{Daly}(1998)}]{Paschmann98}%
	\BibitemOpen
	\bibfield  {author} {\bibinfo {author} {\bibfnamefont {G.}~\bibnamefont
			{Paschmann}}\ and\ \bibinfo {author} {\bibfnamefont {W.~D.}\ \bibnamefont
			{Daly}},\ }\href@noop {} {\emph {\bibinfo {title} {Analysis Methods for
				Multi-Spacecraft Data, ISSI Sci. Rep. SR-001}}}\ (\bibinfo  {publisher} {The
		International Space Science Institute},\ \bibinfo {year} {1998})\BibitemShut
	{NoStop}%
	\bibitem [{\citenamefont {Fu}\ \emph {et~al.}(2015)\citenamefont {Fu},
		\citenamefont {Vaivads}, \citenamefont {Khotyaintsev}, \citenamefont
		{Olshevsky}, \citenamefont {Andr\'e}, \citenamefont {Cao}, \citenamefont
		{Huang}, \citenamefont {Retin\`o},\ and\ \citenamefont {Lapenta}}]{Fu15}%
	\BibitemOpen
	\bibfield  {author} {\bibinfo {author} {\bibfnamefont {H.~S.}\ \bibnamefont
			{Fu}}, \bibinfo {author} {\bibfnamefont {A.}~\bibnamefont {Vaivads}},
		\bibinfo {author} {\bibfnamefont {Y.~V.}\ \bibnamefont {Khotyaintsev}},
		\bibinfo {author} {\bibfnamefont {V.}~\bibnamefont {Olshevsky}}, \bibinfo
		{author} {\bibfnamefont {M.}~\bibnamefont {Andr\'e}}, \bibinfo {author}
		{\bibfnamefont {J.~B.}\ \bibnamefont {Cao}}, \bibinfo {author} {\bibfnamefont
			{S.~Y.}\ \bibnamefont {Huang}}, \bibinfo {author} {\bibfnamefont
			{A.}~\bibnamefont {Retin\`o}}, \ and\ \bibinfo {author} {\bibfnamefont
			{G.}~\bibnamefont {Lapenta}},\ }\href {\doibase 10.1002/2015JA021082}
	{\bibfield  {journal} {\bibinfo  {journal} {Journal of Geophysical Research:
				Space Physics}\ }\textbf {\bibinfo {volume} {120}},\ \bibinfo {pages} {3758}
		(\bibinfo {year} {2015})}\BibitemShut {NoStop}%
	\bibitem [{\citenamefont {Karimabadi}\ \emph {et~al.}(2007)\citenamefont
		{Karimabadi}, \citenamefont {Daughton},\ and\ \citenamefont
		{Scudder}}]{Karimabadi07}%
	\BibitemOpen
	\bibfield  {author} {\bibinfo {author} {\bibfnamefont {H.}~\bibnamefont
			{Karimabadi}}, \bibinfo {author} {\bibfnamefont {W.}~\bibnamefont
			{Daughton}}, \ and\ \bibinfo {author} {\bibfnamefont {J.}~\bibnamefont
			{Scudder}},\ }\href {http://dx.doi.org/10.1029/2007GL030306} {\bibfield
		{journal} {\bibinfo  {journal} {Geophysical Research Letters}\ }\textbf
		{\bibinfo {volume} {34}} (\bibinfo {year} {2007})}\BibitemShut {NoStop}%
	\bibitem [{\citenamefont {Egedal}\ \emph {et~al.}(2011)\citenamefont {Egedal},
		\citenamefont {Le}, \citenamefont {Pritchett},\ and\ \citenamefont
		{Daughton}}]{Egedal11}%
	\BibitemOpen
	\bibfield  {author} {\bibinfo {author} {\bibfnamefont {J.}~\bibnamefont
			{Egedal}}, \bibinfo {author} {\bibfnamefont {A.}~\bibnamefont {Le}}, \bibinfo
		{author} {\bibfnamefont {P.~L.}\ \bibnamefont {Pritchett}}, \ and\ \bibinfo
		{author} {\bibfnamefont {W.}~\bibnamefont {Daughton}},\ }\href {\doibase
		10.1063/1.3646316} {\bibfield  {journal} {\bibinfo  {journal} {Physics of
				Plasmas}\ }\textbf {\bibinfo {volume} {18}},\ \bibinfo {pages} {102901}
		(\bibinfo {year} {2011})}\BibitemShut {NoStop}%
	\bibitem [{\citenamefont {Hwang}\ \emph {et~al.}(2017)\citenamefont {Hwang},
		\citenamefont {Sibeck}, \citenamefont {Choi}, \citenamefont {Chen},
		\citenamefont {Ergun}, \citenamefont {Khotyaintsev}, \citenamefont {Giles},
		\citenamefont {Pollock}, \citenamefont {Gershman}, \citenamefont {Dorelli},
		\citenamefont {Avanov}, \citenamefont {Paterson}, \citenamefont {Burch},
		\citenamefont {Russell}, \citenamefont {Strangeway},\ and\ \citenamefont
		{Torbert}}]{Hwang17}%
	\BibitemOpen
	\bibfield  {author} {\bibinfo {author} {\bibfnamefont {K.-J.}\ \bibnamefont
			{Hwang}}, \bibinfo {author} {\bibfnamefont {D.~G.}\ \bibnamefont {Sibeck}},
		\bibinfo {author} {\bibfnamefont {E.}~\bibnamefont {Choi}}, \bibinfo {author}
		{\bibfnamefont {L.-J.}\ \bibnamefont {Chen}}, \bibinfo {author}
		{\bibfnamefont {R.~E.}\ \bibnamefont {Ergun}}, \bibinfo {author}
		{\bibfnamefont {Y.}~\bibnamefont {Khotyaintsev}}, \bibinfo {author}
		{\bibfnamefont {B.~L.}\ \bibnamefont {Giles}}, \bibinfo {author}
		{\bibfnamefont {C.~J.}\ \bibnamefont {Pollock}}, \bibinfo {author}
		{\bibfnamefont {D.}~\bibnamefont {Gershman}}, \bibinfo {author}
		{\bibfnamefont {J.~C.}\ \bibnamefont {Dorelli}}, \bibinfo {author}
		{\bibfnamefont {L.}~\bibnamefont {Avanov}}, \bibinfo {author} {\bibfnamefont
			{W.~R.}\ \bibnamefont {Paterson}}, \bibinfo {author} {\bibfnamefont {J.~L.}\
			\bibnamefont {Burch}}, \bibinfo {author} {\bibfnamefont {C.~T.}\ \bibnamefont
			{Russell}}, \bibinfo {author} {\bibfnamefont {R.~J.}\ \bibnamefont
			{Strangeway}}, \ and\ \bibinfo {author} {\bibfnamefont {R.~B.}\ \bibnamefont
			{Torbert}},\ }\href {\doibase 10.1002/2017GL072830} {\bibfield  {journal}
		{\bibinfo  {journal} {Geophysical Research Letters}\ }\textbf {\bibinfo
			{volume} {44}},\ \bibinfo {pages} {2049} (\bibinfo {year}
		{2017})}\BibitemShut {NoStop}%
\end{thebibliography}

%
 
\end{document}


\title{Supplemental Material for ``In situ spacecraft observations of a structured electron diffusion region during magnetopause reconnection''}
\author{Giulia Cozzani}
\email{giulia.cozzani@lpp.polytechnique.fr}
\affiliation{Laboratoire de Physique des Plasmas,CNRS/Ecole Polytechnique/Sorbonne Universit\'e, Universit\'e Paris Sud, Observatoire de Paris, Paris, France}
\affiliation{Dipartimento di Fisica ''E. Fermi'', Universit\`a Pisa, Pisa, Italy}
\author{A. Retin\`o}
\affiliation{Laboratoire de Physique des Plasmas,CNRS/Ecole Polytechnique/Sorbonne Universit\'e, Universit\'e Paris Sud, Observatoire de Paris, Paris, France}
\author{F. Califano}
\affiliation{Dipartimento di Fisica ''E. Fermi'', Universit\`a Pisa, Pisa, Italy}
\author{A. Alexandrova}
\affiliation{Laboratoire de Physique des Plasmas,CNRS/Ecole Polytechnique/Sorbonne Universit\'e, Universit\'e Paris Sud, Observatoire de Paris, Paris, France}
\author{O. Le~Contel}
\affiliation{Laboratoire de Physique des Plasmas,CNRS/Ecole Polytechnique/Sorbonne Universit\'e, Universit\'e Paris Sud, Observatoire de Paris, Paris, France}
\author{Y. Khotyaintsev}
\affiliation{Swedish Institute of Space Physics, Uppsala, Sweden}
\author{A. Vaivads}
\affiliation{Swedish Institute of Space Physics, Uppsala, Sweden}
\author{H. S. Fu}
\affiliation{School of Space and Environment, Beihang University, Beijing, China}
\author{F. Catapano}
\affiliation{Dipartimento di Fisica, Universit\`a della Calabria, Rende, Italy}
\affiliation{Laboratoire de Physique des Plasmas,CNRS/Ecole Polytechnique/Sorbonne Universit\'e, Universit\'e Paris Sud, Observatoire de Paris, Paris, France}
\author{H. Breuillard}
\affiliation{Laboratoire de Physique des Plasmas,CNRS/Ecole Polytechnique/Sorbonne Universit\'e, Universit\'e Paris Sud, Observatoire de Paris, Paris, France}
\affiliation{Laboratoire de Physique et Chimie de l'Environnement et de l'Espace, CNRS-Universit\'e d'Orl\'eans, France}
\author{N. Ahmadi}
\affiliation{Laboratory of Atmospheric and Space Physics, University of Colorado Boulder, Boulder, Colorado, USA}
\author{P.-A. Lindqvist}
\affiliation{KTH Royal Institute of Technology, Stockholm, Sweden}
%
%
\author{R. E. Ergun}
\affiliation{Laboratory of Atmospheric and Space Physics, University of Colorado Boulder, Boulder, Colorado, USA}
\author{R. B. Torbert}
\affiliation{Space Science Center, University of New Hampshire, Durham, New Hampshire, USA}
\author{B. L. Giles}
\affiliation{NASA Goddard Space Flight Center, Greenbelt, Maryland, USA}
\author{C. T. Russell}
\affiliation{Department of Earth and Space Sciences, University of California, Los Angeles, California, USA}

\author{R. Nakamura}
\affiliation{Space Research Institute, Austrian Academy of Sciences, Graz, Austria}
\author{S. Fuselier}
\affiliation{Southwest Research Institute, San Antonio, Texas, USA}
\author{B. H. Mauk}
\affiliation{The Johns Hopkins University Applied Physics Laboratory, Laurel, Maryland, USA}
\author{T. Moore}
\affiliation{NASA Goddard Space Flight Center, Greenbelt, Maryland, USA}
\author{J. L. Burch}
\affiliation{Southwest Research Institute, San Antonio, Texas, USA}
\begin{abstract}
In this Supplemental Material we provide additional information on the computation of the energy conversion between the fields and the particles $\mathbf{E}' \cdot \mathbf{J}$ and its associated error. We show that the uncertainties can be of the same order of the measured quantities, preventing the estimation of the dissipation on all the probes. 
\end{abstract}


\maketitle

%


\section{$\mathbf{E}' \cdot \mathbf{J}$ and the associated error for the four spacecraft}

We here provide the computation of the maximum errors on $\mathbf{E}' \cdot \mathbf{J}$, $E'_L J_L$, $E'_M J_M$, $E'_N J_N$. The maximum error on $E'_i J_i$ is computed as follows
\begin{equation}
\delta (E'_i J_i) = J_i \ \delta E'_i + E'_i \ \delta J_i 
\end{equation}

in which all the quantities are supposed to be not correlated and $i = L,M,N$. The error on the electric field is $\delta E_i = 20 \% |E|$ when $|E| > 1 \ mV/m$ and $\delta E_i = 0.1 \ mV/m$ otherwise, based on statistical analysis on $\mathbf{E}$ and $\delta E_i$. The errors on the electron and ions moments are $\sim 10 \%$ and the magnetic field error is $\delta B = 0.5 \ nT$. The error on  $\mathbf{E}' \cdot \mathbf{J}$ is  $ \delta (\mathbf{E}' \cdot \mathbf{J}) = \sum_{i = L,M,N} \delta (E'_i J_i)$. We notice that the behavior of the different dissipation term is qualitatively the same for all the spacecraft, for example $E'_N J_N$ and $E'_M J_M$ is bipolar for all the probes (see Fig.1). Nevertheless, the errors are comparable to the measured quantities except for MMS4, the only satellite for which the negative and positive peaks of the dissipation terms are beyond the errors.

\begin{figure}
	\centering
	\includegraphics[width=0.8\columnwidth]{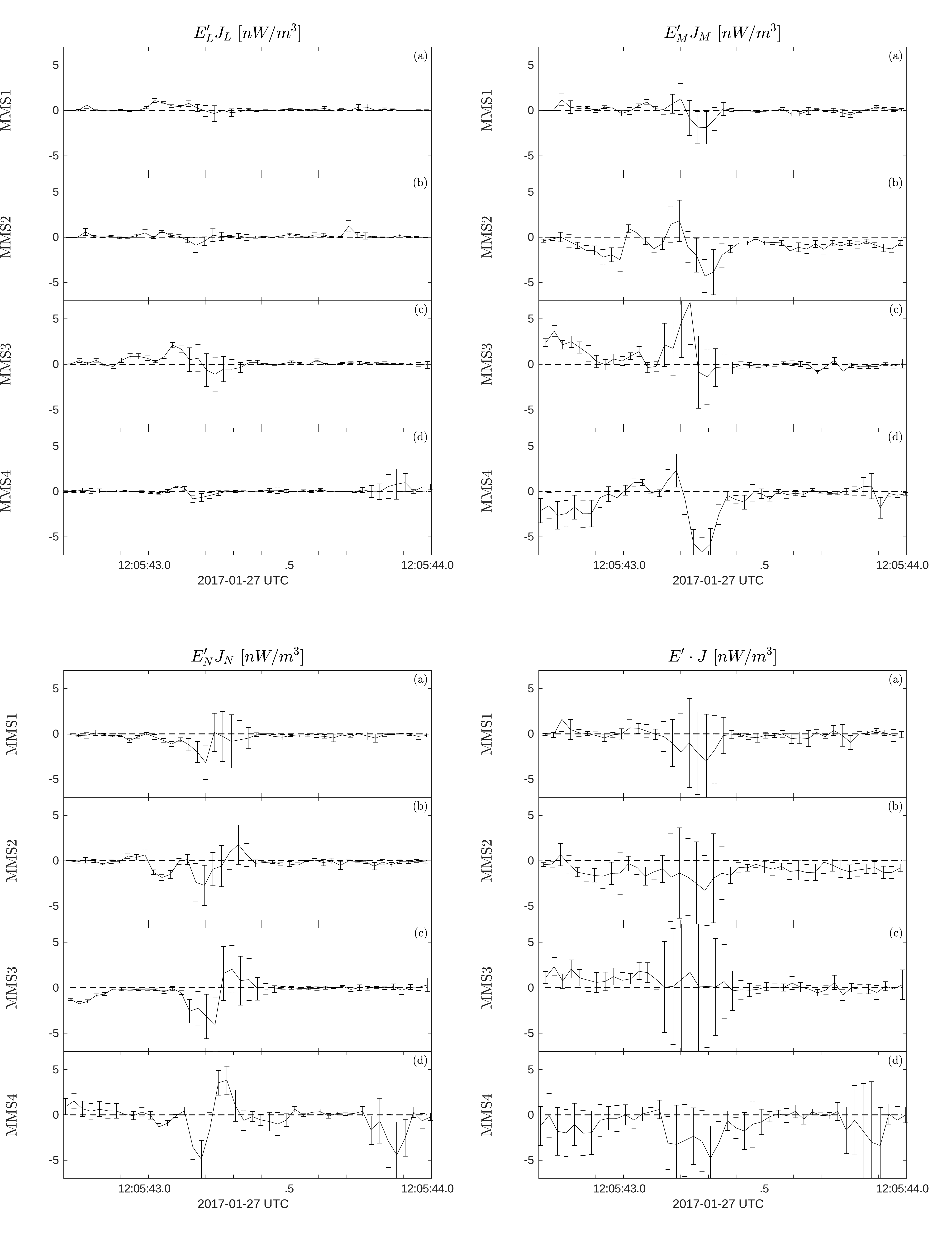}
	\caption{\textbf{Top left}: $E'_L J_L$ and its associated maximum error for the four spacecraft. \textbf{Top right}: $E'_M J_M$ and its associated maximum error for the four spacecraft. \textbf{Bottom left}: $E'_N J_N$ and its associated maximum error for the four spacecraft. \textbf{Bottom right}: $\mathbf{E}' \cdot \mathbf{J}$ and its associated maximum error for the four spacecraft.\label{fig1}}
\end{figure}